\documentclass[lettersize,journal]{IEEEtran}
\usepackage{amsmath,amsfonts}
\usepackage{algorithmic}
\usepackage{algorithm}
\usepackage{array}
\usepackage[caption=false,font=normalsize,labelfont=sf,textfont=sf]{subfig}
\usepackage{textcomp}
\usepackage{stfloats}
\usepackage{url}
\usepackage{verbatim}
\usepackage{graphicx}
\usepackage{booktabs}
\usepackage{cite}

\usepackage{hyperref}
\usepackage{tabularx}
\usepackage{pifont}
\usepackage{siunitx}
\usepackage{stackengine}
\stackMath
\newcolumntype{Y}{>{\centering\arraybackslash}X}

\begin{document}

\title{ThermoDSE: A \underline{Therm}al‑Aware and C\underline{o}mprehensive \underline{D}esign \underline{S}pace \underline{E}xploration for Chiplet‑Based DNN Accelerators}

\author{Jian Peng, \emph{Student Member, IEEE}, Hanwei Fan, \emph{Student Member, IEEE}, Jingbo Jiang, \emph{Student \\Member, IEEE}, Lin Jiang, \emph{Member, IEEE}, Wei Zhang, \emph{Fellow, IEEE}
\thanks{Manuscript received XXXX; revised XXXX; accepted XXXX. Data of publication XXXX; Date of current version XXXX. This work is partially funded by Hong Kong RGC GRF XXXX. Corresponding author: Wei Zhang.}
\thanks{Jian Peng (jpengai@connect.ust.hk), Hanwei Fan (hfanah@connect.ust.hk), Jingbo Jiang (jjiangan@connect.ust.hk), and Wei Zhang (wei.zhang@ust.hk) are with the Dept. of Electronic and Computer Engineering, Hong Kong University of Science and Technology.}
\thanks{Lin Jiang (jianglin1@neu.edu.cn) is with Northeastern University, Shenyang, China.}
}

\markboth{Journal of \LaTeX\ Class Files,~Vol.~14, No.~8, August~2021}%
{Shell \MakeLowercase{\textit{et al.}}: A Sample Article Using IEEEtran.cls for IEEE Journals}


\maketitle

\begin{abstract}
Chiplet‑based DNN accelerators provide a scalable path to balance performance and yield for modern AI workloads. However, such systems face critical challenges in area and thermal constraints. The design space optimization should jointly consider the fine-grained task modeling, chiplet granularity, core granularity, and critical physical constraints. To the best of our knowledge, we are the first framework that involves all these factors. 
In this work, we propose ThermoDSE, a thermal‑aware and comprehensive design space exploration framework for chiplet‑based DNN accelerators. ThermoDSE integrates existing fine‑grained modeling techniques into a uniform simulation and optimization framework that jointly considers architecture design, task orchestration, and inter‑chiplet communication under strict thermal and area constraints. Experimental results show that ThermoDSE achieves up to 3.5× improvement in Energy-Delay-Inverse-Yield (defined as $E\times D \times Y^{-1}$) cost, compared with state‑of‑the‑art Simba and other baselines. Furthermore, relative to simulated annealing and reinforcement learning–based methods, ThermoDSE convert to better design points with 3.7× and 29.4× running time speed up, respectively.
\end{abstract}


\section{Introduction}
\IEEEPARstart{C}{hiplet} technology has recently attracted significant attention from both industry and academia due to the slowing of Moore's Law. Smaller chiplets benefit from higher fabrication yields and shorter design cycles. By leveraging advanced packaging technologies, multiple chiplets can be integrated into a single package, delivering a flexible, high-performance processor at a reasonable cost. 
As computing demand for machine learning continues to increase, DNN accelerators are designed to meet stringent requirements, including high throughput, energy efficiency, and reliability. DNN accelerators featuring arrays of processing elements (PEs) have emerged as an efficient platform and have steadily grown in scale. 
For example, Cerebras~\cite{cerebras2025} employs a wafer-scale accelerator to deliver cluster-scale throughput. However, such large monolithic dies result in significantly higher manufacturing cost. Moreover, due to their complexity, these accelerators face long development cycles and high non-recurring engineering (NRE) costs~\cite{feng2022chiplet}. 

Chiplet-based DNN accelerators provide a promising solution to these problems. Simba~\cite{shao2019simba}, proposed by NVIDIA in 2019, is a 36-chiplet prototype for deep-learning inference. Each chiplet is an NVDLA-like NPU core that achieves a 4-TOPS peak performance within an area of only $2.5\,mm \times2.4\,mm$, leading to lower manufacturing and NRE costs. Several advantages arise from splitting a large monolithic DNN accelerator into smaller chiplets. 
First, smaller dies achieve much higher fabrication yield. Second, scalability improves since the number of chiplets can be easily extended with shorter development cycles. 

For chiplet-based DNN accelerators, core communication—including the Network-on-Chip (NoC) for intra-chiplet data movement and the Network-on-Package (NoP) for die-to-die communication—provides significant benefits in fine-grained data parallelism and data reuse during workload orchestration. 
Several previous studies have focused on dataflow and task-mapping optimization for chiplet-based DNN accelerators. 
Tangram~\cite{gao2019tangram} and SET~\cite{cai2023inter} explore layer-pipelined execution to optimize dataflow. 
Aotmic~\cite{zheng2022atomic} and Klotski~\cite{bai2024klotski} propose graph-level workload orchestration methods. 
Through precise graph partitioning, scheduling, and mapping, they achieve substantial improvements in performance, hardware utilization, and energy efficiency. 
Some previous works~\cite{tan2021nn
, cai2024gemini} jointly explore workload orchestration and chiplet design. 
However, these pioneering works primarily focus on novel encodings of workload orchestration for task scheduling and mapping. Their design space exploration (DSE) of chiplet- and accelerator-level architecture remains limited and incomplete. 
Chiplet-Gym~\cite{mishty2024chiplet} introduces RL for chiplet optimization and considers comprehensive package-level factors such as yield, packaging cost, and interconnect bandwidth. 
Nevertheless, it relies on a simple analytical model that ignores chiplet granularity exploration, core-level optimization, and the data reuse enabled by fine-grained modeling.

On the other hand, chiplet-based DNN accelerators suffer from severe thermal-wall problems due to higher integration density. These issues should be evaluated and addressed in the early design stage. Unfortunately, none of the above studies considers these critical constraints. 
Mathur et al.~\cite{mathur2021thermal} propose a thermal-aware framework for 3D DNN accelerator exploration, but it supports only a single-core accelerator. 
TESA~\cite{shukla2023temperature} introduces a temperature-aware methodology that adjusts the sizes and inter-chiplet spacing (ICS) of accelerators to jointly balance cost and performance under area and thermal constraints. 
However, the NPU cores in TESA operate independently without NoC/NoP to simplify the simulation, thereby ignoring core-level optimization and the data reuse benefits of inter-core communication.

Therefore, we argue that a comprehensive optimization framework for chiplet-based DNN accelerators should incorporate the following aspects. 
\textbf{Fine-grained task modeling}: 
Due to the presence of NoC/NoP, fine-grained task orchestration can improve performance by over $40\%$ compared with coarse-grained layer-wise scheduling~\cite{zheng2022atomic}. 
Moreover, effective task orchestration enables accurate DSE results and precise thermal simulation. 
\textbf{Considering multiple constraints}: 
During optimization, several constraints must be strictly satisfied. 
For chiplet designs, both peak temperature and interposer area are tightly regulated. 
\textbf{Effective DSE algorithm}: 
The design space of chiplet-based DNN accelerators is enormous, covering both core-level and chiplet-level architectural parameters. 
Furthermore, because fine-grained task modeling is time-consuming, the DSE algorithm must efficiently locate near-optimal design points with limited iterations.

\begin{figure*}[!t]
\centering
\subfloat[]{\includegraphics[width=0.5 \linewidth]{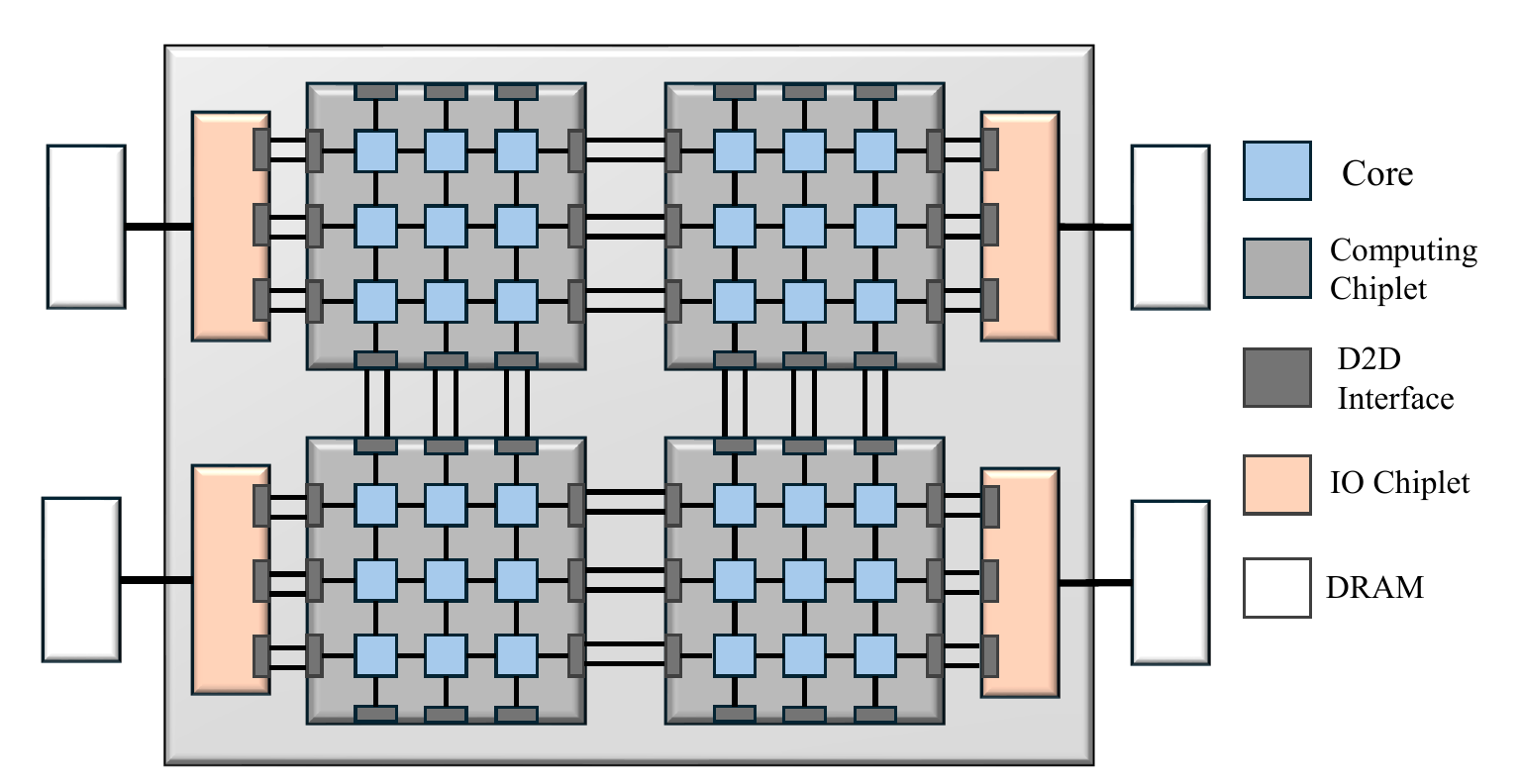}%
\label{fig_smc_arch}}
\hfil
\subfloat[]{\includegraphics[width=0.34 \linewidth]{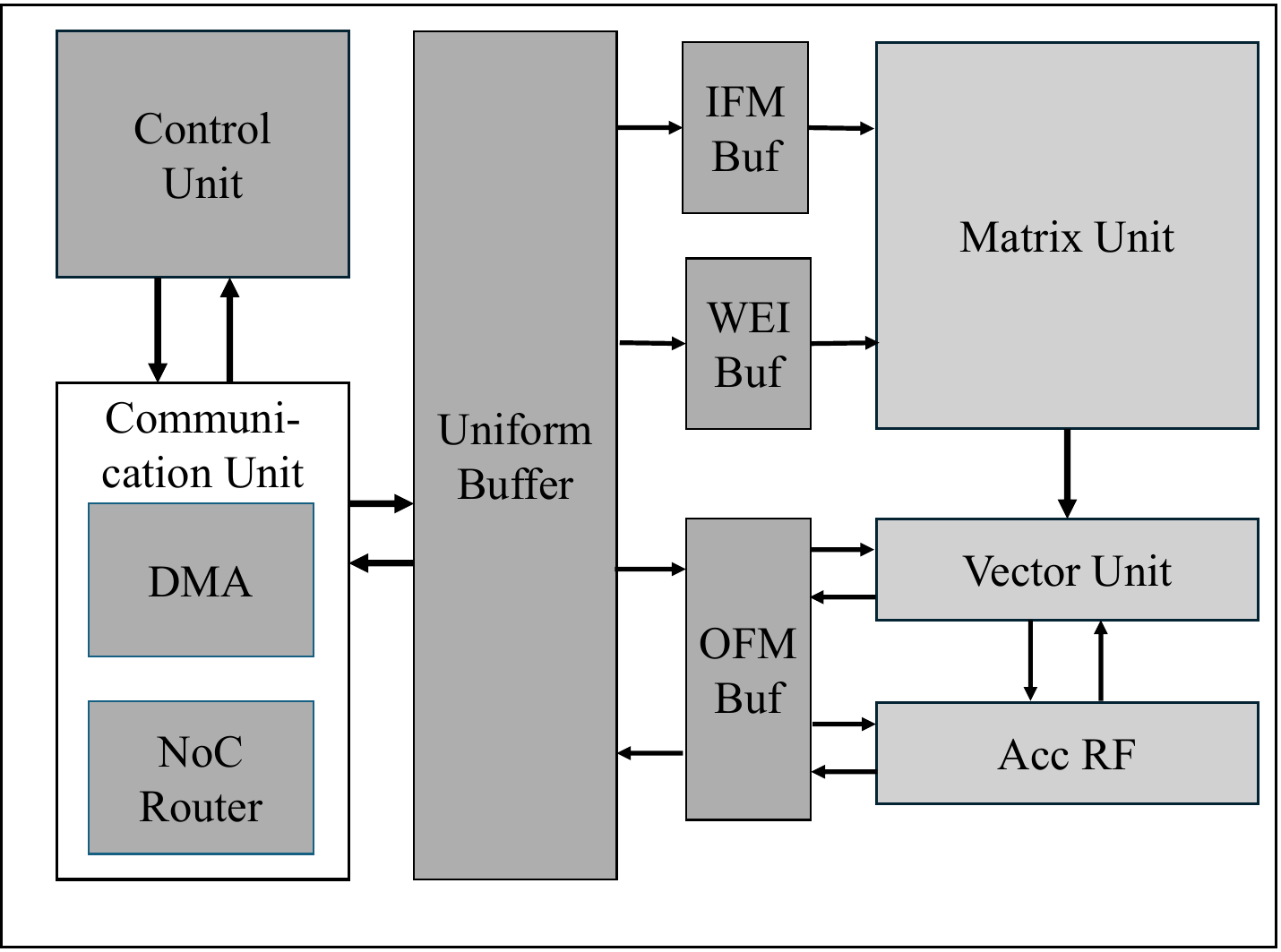}%
\label{fig_core_arch}}
\hfil
\caption{(a) Diagram of a chiplet-based DNN accelerator, and (b) Architecture of a single NPU core.}
\label{fig_acc_arch}
\vspace{-2mm}
\end{figure*}

To bridge these gaps, we develop \textbf{ThermoDSE}, a thermal-aware, comprehensive, and efficient DSE framework for chiplet-based DNN accelerators. 
We construct a thermal and performance simulator with fine-grained task orchestration to deliver high-performance and accurate simulation results. 
An effective DSE algorithm based on transductive experimental design (TED)~\cite{liu2013learning} and scalable constrained Bayesian Optimization (SCBO)~\cite{eriksson2021scalable} is then applied to optimize under multiple constraints. 
Our main contributions are summarized as follows:
\begin{itemize}
    \item We build a comprehensive performance and thermal simulation framework that integrates fine-grained task modeling, including $\mu$task-level data reuse and buffering. The simulator provides accurate and efficient results for iterative DSE progress.
    \item We propose an effective DSE algorithm based on TED and SCBO. The TED-based initialization generates a well-distributed initial sample set across the design space, retaining essential information for surrogate modeling. Meanwhile, the SCBO algorithm extends Bayesian optimization for multi-constraint scenarios by dynamically adjusting the trust region during each iteration, enabling efficient search around boundary conditions. 
    \item Finally, we validate our methodology by comparing the Energy-Delay-Inverse-Yield (EDYP) cost against industrial product Simba and prior frameworks (TESA and Chiplet-Gym). Experimental results show that our framework achieves up to 3.5x lower EDYP than these baselines. We further demonstrate that the proposed DSE method attains better design points with 3.7x and 29.4x speedup over conventional simulated annealing (SA) and reinforcement learning (RL) based methods, respectively.
\end{itemize}

The rest of this paper is organized as follows. Section~II introduces the background of chiplet-based DNN accelerator design and presents a motivating example. 
Section~III reviews related works. 
Section~IV describes the proposed methodology, including the overall framework, fine-grained task orchestration, modeling, and DSE formulation. 
Section~V presents experimental results and analysis. 
Finally, Section~VI concludes the paper and discusses future work.

\section{Background and Motivation}
\subsection{Chiplet-based Scalable DNN Accelerator}

The chiplet-based scalable DNN accelerator was first introduced by NVIDIA's Simba~\cite{shao2019simba} and has been extensively explored in prior studies~\cite{gao2019tangram, tan2021nn, lai2025optimizing, cai2024gemini}. The overall architecture is shown in Fig.~\ref{fig_acc_arch}. The main component of this architecture is a computing NPU core array. Dividing a large monolithic core array into multiple computing chiplets can significantly reduce both NRE and manufacturing costs. A mesh-based NoC interconnects all computing cores within a chiplet as well as the controller within the I/O chiplets, enabling arbitrary core-to-core, core-to-DRAM, and DRAM-to-core communications. The Network-on-Package (NoP) supports inter-chiplet communication. The die-to-die (D2D) transmitter (TX) within a chiplet encodes the data independently and forwards it to the corresponding D2D receiver (RX) in another chiplet. The D2D RX decodes the incoming data and proceeds with NoC transmission. This inter-chiplet communication is fully automatic and transparent to both the source and destination cores. The I/O chiplet contains an array of I/O modules that interface with DRAM, host systems, or external data sources such as cameras. 

Fig.~\ref{fig_acc_arch}(b) illustrates the architecture of a single NPU core. Each core typically comprises four main components: a control unit, a communication unit, buffers, a matrix unit, and a vector unit. The control unit manages the operation of the NPU core according to the received instructions. The communication unit handles both inter-core and intra-core data movement through the NoC router and DMA engine, respectively. The uniform buffer (\textit{ubuf}) usually has a large capacity and can store all types of data. The IFM and WEI buffers (\textit{ibuf} and \textit{wbuf}) store input feature maps and weight parameters, respectively. Partial sums and output feature maps (OFMs) are stored in the accumulator register file (\textit{Acc RF}) and OFM buffer, respectively. The matrix unit typically contains a systolic array to perform convolution and matrix multiplication operations, while the vector unit executes layer-specific functions such as activation, pooling, and normalization.

\begin{figure}[!t]
\centering
    \subfloat[GoogLeNet]{\scriptsize \includegraphics[width=0.29 \linewidth]{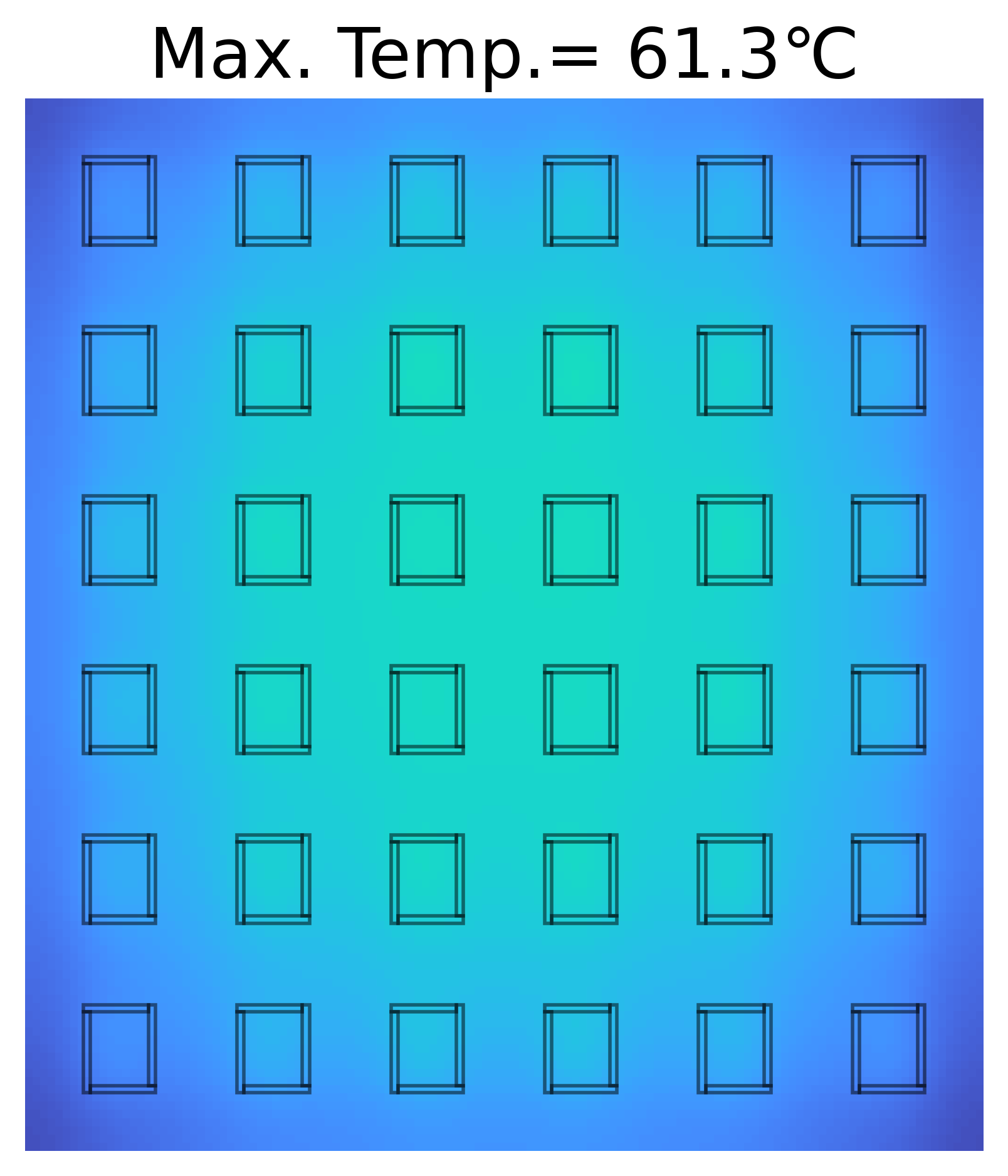}%
\label{fig_resnet_temp}}
\hfil
\subfloat[BERT]{\includegraphics[width=0.29 \linewidth]{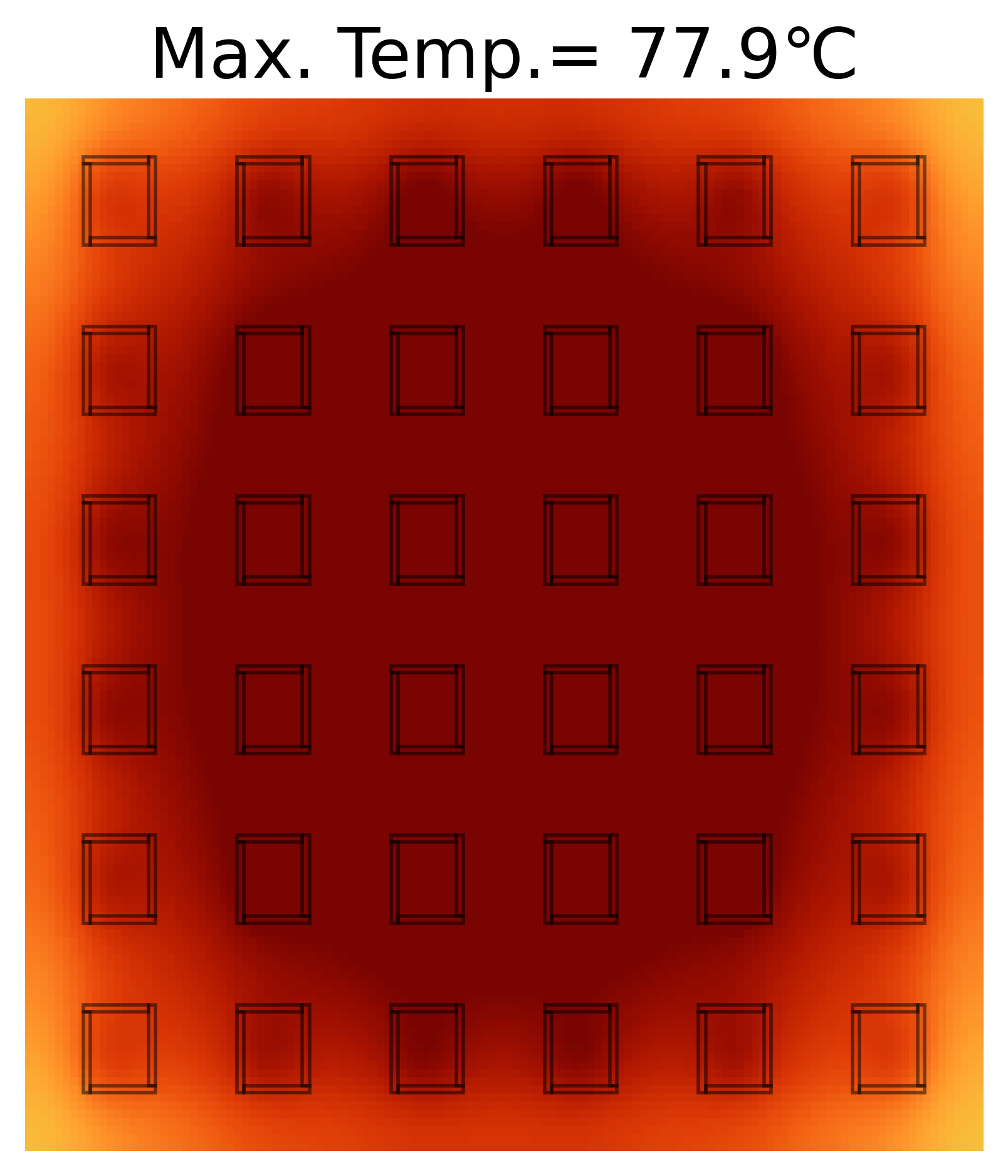}%
\label{fig_transformer_temp}}
\hfil
\subfloat[AR/VR\ \ \ \ \ \ \ \ \ \ \ \ \ \ ]{\includegraphics[width=0.408 \linewidth]{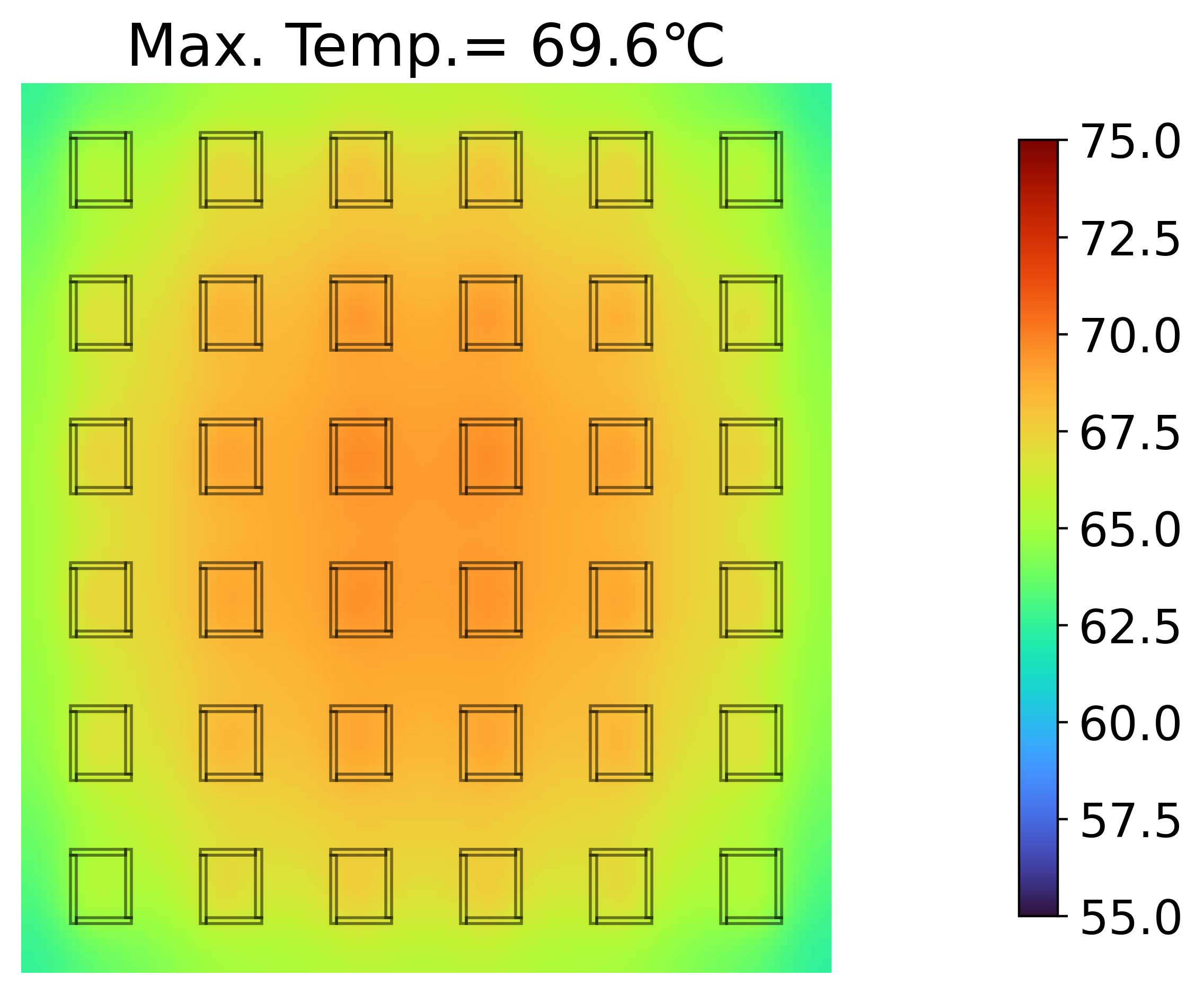}%
\label{fig_AR_temp}}
\hfil
\caption{Thermal distribution of three different workloads on the Simba prototype: (a) GoogLeNet, with a peak temperature of \SI{61.3}{\degreeCelsius}; (b) Transformer, with a peak temperature of \SI{77.9}{\degreeCelsius}; and (c) AR/VR workloads composed of six types of DNNs, with a peak temperature of \SI{69.6}{\degreeCelsius}.}
\label{fig_simba_temp}
\vspace{-2mm}
\end{figure}

\subsection{Multi-DNN Workload}

The success of deep learning in recent years has led to breakthrough applications such as augmented and virtual reality (AR/VR)~\cite{wu2019machine} and autonomous driving~\cite{tian2018deeptest, lu2025taco, kong2024wts}. These applications typically employ multiple DNNs to perform distinct tasks collaboratively, achieving state-of-the-art system performance. For example, AR/VR applications involve several types of neural networks, such as image segmentation (U-Net), object detection (MobileNet and Yolo), object recognition (ResNet-50 and GoogLeNet), and speech recognition (BERT). These networks are executed sequentially with small batch sizes~\cite{kwon2021heterogeneous}. 

However, different types of networks exhibit widely varying power and thermal characteristics. First, models such as BERT and Yolo demand significantly higher computational power than conventional CNNs. Previous studies~\cite{wu2019accelergy, peng2024data} demonstrated that the sparsity of input feature maps greatly affects matrix-unit power consumption, owing to the zero-bypass mechanism in MAC units. Meanwhile, activation functions such as GLUE in Transformers and LeakyReLU in Yolo rarely induce sparsity compared with ReLU in conventional networks~\cite{pao2025eda}. Second, the attention layers in BERT involve intensive data movement, leading to considerably higher power consumption in communication units. This diversity in DNN characteristics poses substantial challenges for thermal-aware design optimization in chiplet-based DNN accelerators.

\begin{figure}[!t]
\centering
    \subfloat[D1]{\scriptsize \includegraphics[width=0.275 \linewidth]{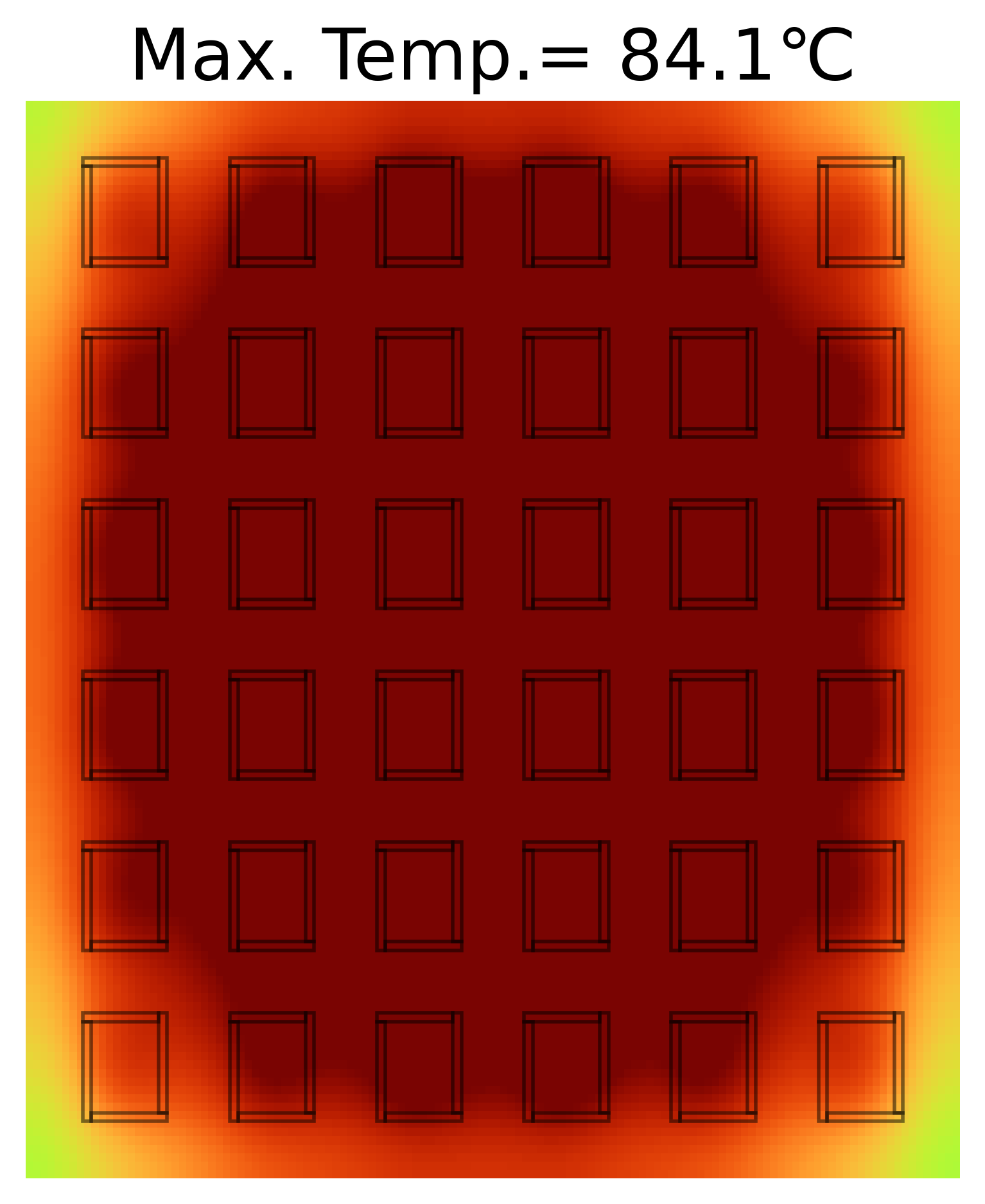}%
\label{fig_simba_L}}
\hfil
\subfloat[D2]{\includegraphics[width=0.275 \linewidth]{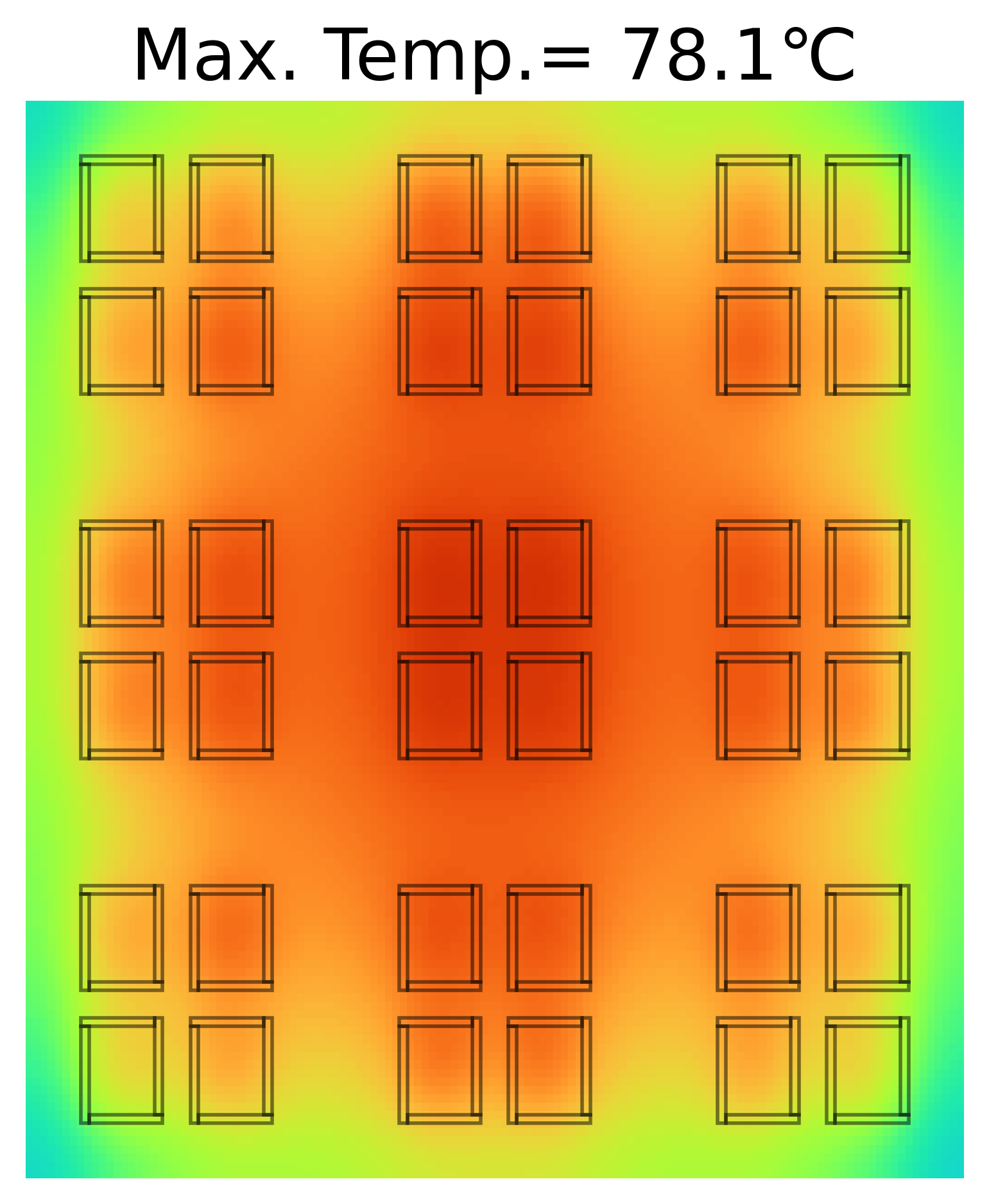}%
\label{fig_simba_L3x3}}
\hfil
\subfloat[D3\ \ \ \ \ \ \ \ \ \ \ \ \ ]{\includegraphics[width=0.38 \linewidth]{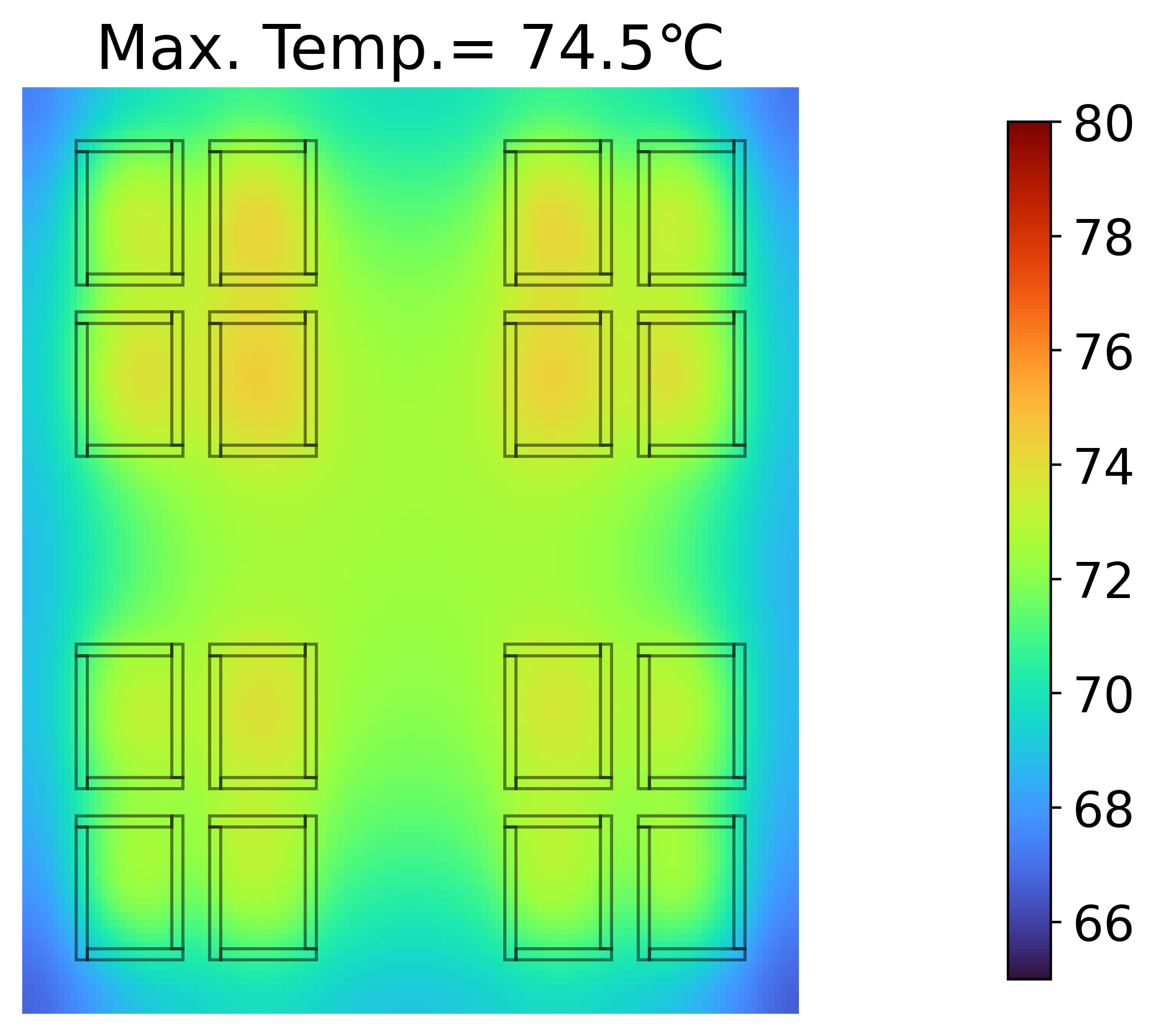}}%

\hfil
\vspace{-1mm}
\caption{Possible directions for miminmizing EDYP cost: (a) D1: allocate more resources to reduce the delay. (b) D2: Increase the chiplet granularity and ICS. (c) D3: Reduce core numbers and allocate more resources to each core.}
\label{fig_mtv_example}
\vspace{-2mm}
\end{figure}

\subsection{Motivations}

Fig.~\ref{fig_simba_temp} illustrates the thermal distribution across three representative workloads on the Simba architecture. We analyze GoogLeNet, BERT, and an AR/VR mixed workload. The lightweight GoogLeNet workload results in a peak temperature of only 61.3\,\textcelsius, whereas BERT produces a peak temperature that is \SI{16.6}{\degreeCelsius} higher, due to its dense input data and higher data movement intensity. For the AR/VR application that integrates multiple DNNs, the peak temperature reaches 69.6\,\textcelsius, falling between the two extremes. Dynamic Voltage and Frequency Scaling~\cite{rahim2025survey} is a classical technique to mitigate power consumption in high-power neural networks during runtime. However, in multi-DNN workloads for mobile or edge scenarios, networks are executed sequentially, making DVFS unsuitable due to its long adjustment latency~\cite{lin2023energy}. 

Obtaining a balanced trade‑off among energy, delay, and yield for chiplet-based designs remains highly challenging in mobile or edge scenarios~\cite{shukla2023temperature,russo2022multiobjective}. Fig.~\ref{fig_mtv_example} presents a motivating example illustrating several potential design directions during DSE. We assume the design of a chiplet-based DNN accelerator is constrained by a peak temperature of \SI{75}{\degreeCelsius} and an interposer area of $300\,mm^2$. In Fig.~\ref{fig_mtv_example}(a), doubling both the computing units and on-chip buffer capacity achieves $1.6\times$ lower delay, but raises the peak temperature to 84.1\,\textcelsius. As shown in Fig.~\ref{fig_mtv_example}(b), reducing the chiplet array from $6\times6$ to $3\times3$ slightly degrades yield by 0.1 but achieves $1.2\times$ lower energy cost and reduces the peak temperature by 6\,\textcelsius. Fewer chiplets reduce interposer-level D2D energy and provide greater ICS, improving cooling efficiency. However, the area constraint limits ICS expansion, and the temperature constraint remains violated. Finally, as shown in Fig.~\ref{fig_mtv_example}(c), reducing the number of NPU cores while increasing the resources per core further lowers communication energy and enlarges ICS, satisfying thermal constraints. Nonetheless, utilization of large systolic arrays drops, slightly increasing delay, and larger SRAMs also raise access energy~\cite{muralimanohar2009cacti}. Therefore, developing a DSE framework that simultaneously considers performance, yield, and physical constraints for chiplet-based DNN accelerators is crucial but non-trivial.

\section{Related Works}

\subsection{Chiplet-Based DNN Accelerator Architecture Exploration}

Recently, several studies have focused on architecture exploration for chiplet-based DNN accelerators. TESA~\cite{shukla2023temperature} is a design exploration framework for temperature-aware sizing of chiplet-based accelerators. It searches for the optimal size of each NPU core and ICS to satisfy temperature, area, and performance constraints through an SA algorithm. Each NPU in TESA operates independently, with its own DRAM channel. However, this architecture is inefficient and impractical for large-scale DNN accelerators because it is unrealistic to allocate so many DRAM channels. Moreover, since TESA only supports network-level parallelism, the utilization of chiplets becomes extremely low when the number of networks is smaller than the number of chiplets.

Chiplet-Gym~\cite{mishty2024chiplet} is an analytical exploration framework that supports different chiplet integration technologies. Although the overall architecture resembles that of Simba, its simple task mapping assumption based on an analytical model cannot capture the detailed characteristics of data reuse and buffer sharing between cores, resulting in inaccurate simulation results. Furthermore, the absence of a detailed single-core model further reduces the accuracy of its performance estimation.

For chiplet-based DNN accelerator design, a number of works have focused on dataflow and task orchestration optimizations~\cite{bai2024klotski, zheng2022atomic, tan2021nn, gao2019tangram, cai2023inter, cai2024gemini}. Unfortunately, except for TESA, most of these studies ignore the critical thermal and area constraints inherent in chiplet-based designs. Moreover, as shown in Fig.~\ref{fig_mtv_example}, the granularity of chiplet partitioning has a significant influence on the thermal profile, yet neither TESA nor Chiplet-Gym explicitly considers the chiplet granularity exploration.

\subsection{DSE Algorithms for Chiplet-Based DNN Accelerators}

NN-Batton~\cite{tan2021nn} and Gemini~\cite{cai2024gemini} search exhaustively for optimal designs due to a limited design space. TESA~\cite{shukla2023temperature} employs the SA algorithm to perform design-space exploration under temperature, area, and performance constraints. However, SA often struggles to obtain high-quality design points because it easily becomes trapped in local optima. To alleviate this issue, Chiplet-Gym~\cite{mishty2024chiplet} introduces RL to guide the search process. Nevertheless, RL converges slowly and is extremely time-consuming for fine-grained models involving large design spaces.

\begin{table}[t]
    \vspace{-2mm}
    \caption{Summary of chiplet-based DNN Accelerator optimizers}
    \centering
    \begin{tabularx}{\linewidth}{c|Y|Y|Y|Y|Y}
    \hline 
    \textbf{Work} & \textbf{Task Grain.} & \textbf{Chiplet Gran.} & \textbf{Core Model} & \textbf{Const.} & \textbf{DSE Method} \\\hline
        ~\cite{tan2021nn} & Fine & \ding{55} & \ding{55} & N.A. & Exhaustive \\ \hline
        ~\cite{shukla2023temperature} & Coarse & \ding{55} & \ding{51} & T and A & SA \\ \hline
        ~\cite{cai2024gemini} & Fine & \ding{51} & \ding{55} & N.A. & Exhaustive \\ \hline 
        ~\cite{mishty2024chiplet} & analytical & \ding{55} & \ding{55} & N.A. & RL\\ \hline \hline
        \textbf{Ours} & Fine & \ding{51} & \ding{51} & T and A & TED+SCBO \\ \hline
        \end{tabularx}
    \label{tab:summary_opt}
    \par
    \textbf{Note:} T and A in Const. column are chip temperature and area, respectively.
\end{table}
To overcome these limitations, we propose \textbf{ThermoDSE}, a therma-aware and comprehensive DSE framework for chiplet-based DNN accelerators. The framework supports multi-level modeling, including chiplet-level granularity, fine-grained task orchestration, NPU-core design, and inter-chiplet spacing. An effective DSE algorithm based on TED and SCBO is employed to explore the design space under thermal and area constraints. SCBO dynamically adjusts the trust region during the optimization process, enabling abundant exploration and accurate identification of optimal design points near constraint boundaries.

Table~\ref{tab:summary_opt} summarizes existing optimization frameworks for chiplet-based DNN accelerators. Compared with prior works, our proposed framework is more comprehensive and effective. Furthermore, the proposed DSE algorithm achieves superior design points with significantly fewer iterations.

\section{Methodology}
\begin{figure}[!t]
\centering
\includegraphics[width=\linewidth]{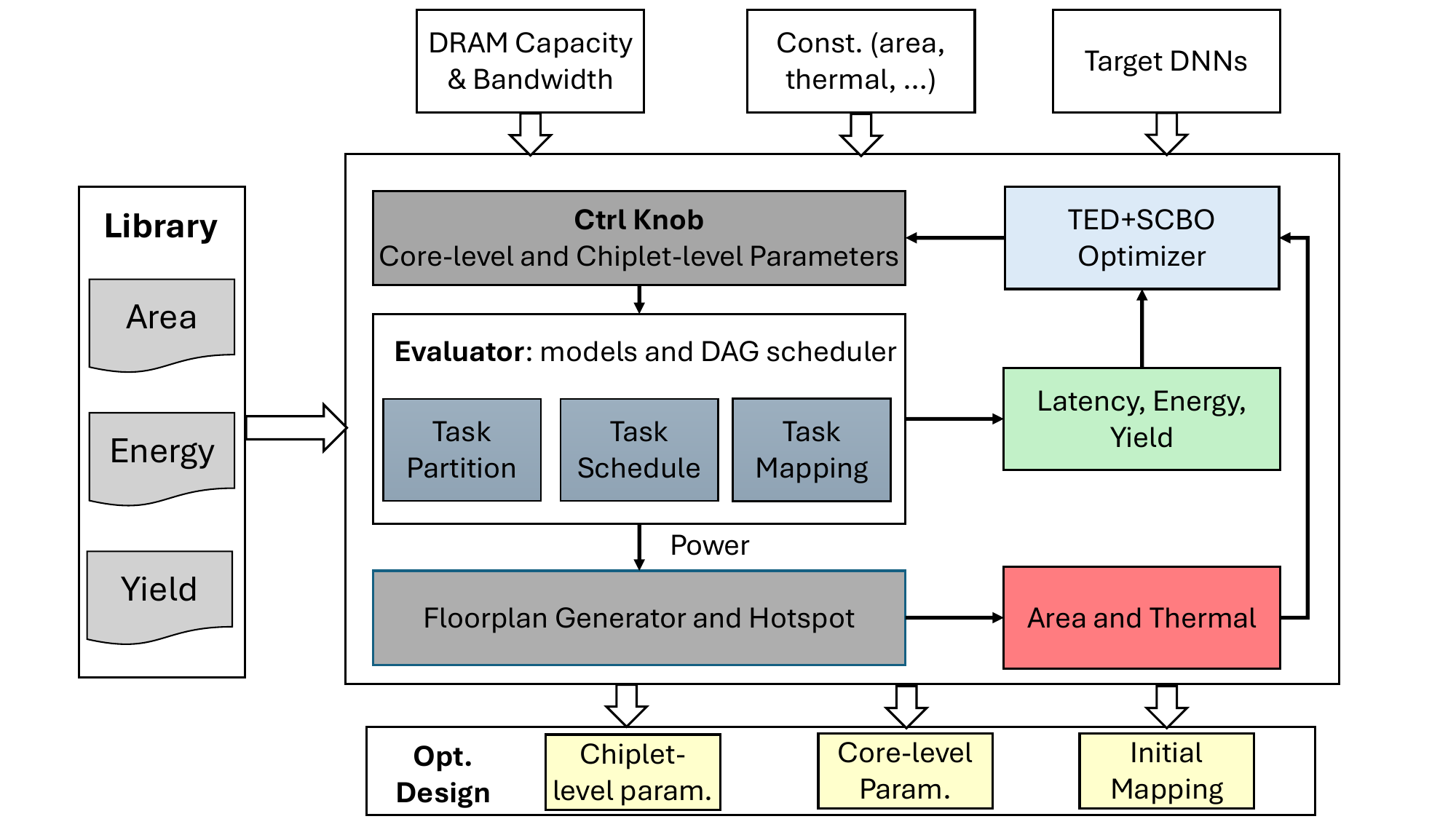}
\caption{The complete framework of design space exploration.}
\label{fig_framework}
\end{figure}

\subsection{Overall Framework}

The overall framework is illustrated in Fig.~\ref{fig_framework}. Our framework aims to obtain the optimal chiplet design for chiplet-based DNN accelerators given the DRAM configuration, design constraints, and target DNN workloads. The final outputs include: (1) core-level DNN accelerator microarchitecture designs, (2) chiplet-level parameters, and (3) an initial thermal-aware task orchestration scheme that satisfies thermal constraint. 

The performance evaluator estimates energy, latency, and yield based on the candidate design, task orchestration, and reference library. To evaluate thermal behavior, we employ HotSpot~6.0~\cite{zhang2015hotspot}, which estimates the steady-state temperature distribution using the provided power map and chiplet floorplan.

\subsection{Fine-grained Task Orchestration}

Task orchestration in chiplet-based DNN accelerators involves complex optimization and has been widely explored in prior works. Among existing methods, graph-based orchestration has attracted significant attention due to its capability to support fine-grained ($\mu$task-level) parallelism, maximize data reuse, and achieve high core utilization~\cite{zheng2022atomic, bai2024klotski}. This approach is particularly effective for small batch sizes (typically $\leq 2$). Therefore, we adopt the graph-based orchestration approach from~\cite{zheng2022atomic} as the foundation of our framework.

\begin{figure}[!t]
\centering
\includegraphics[width= \linewidth]{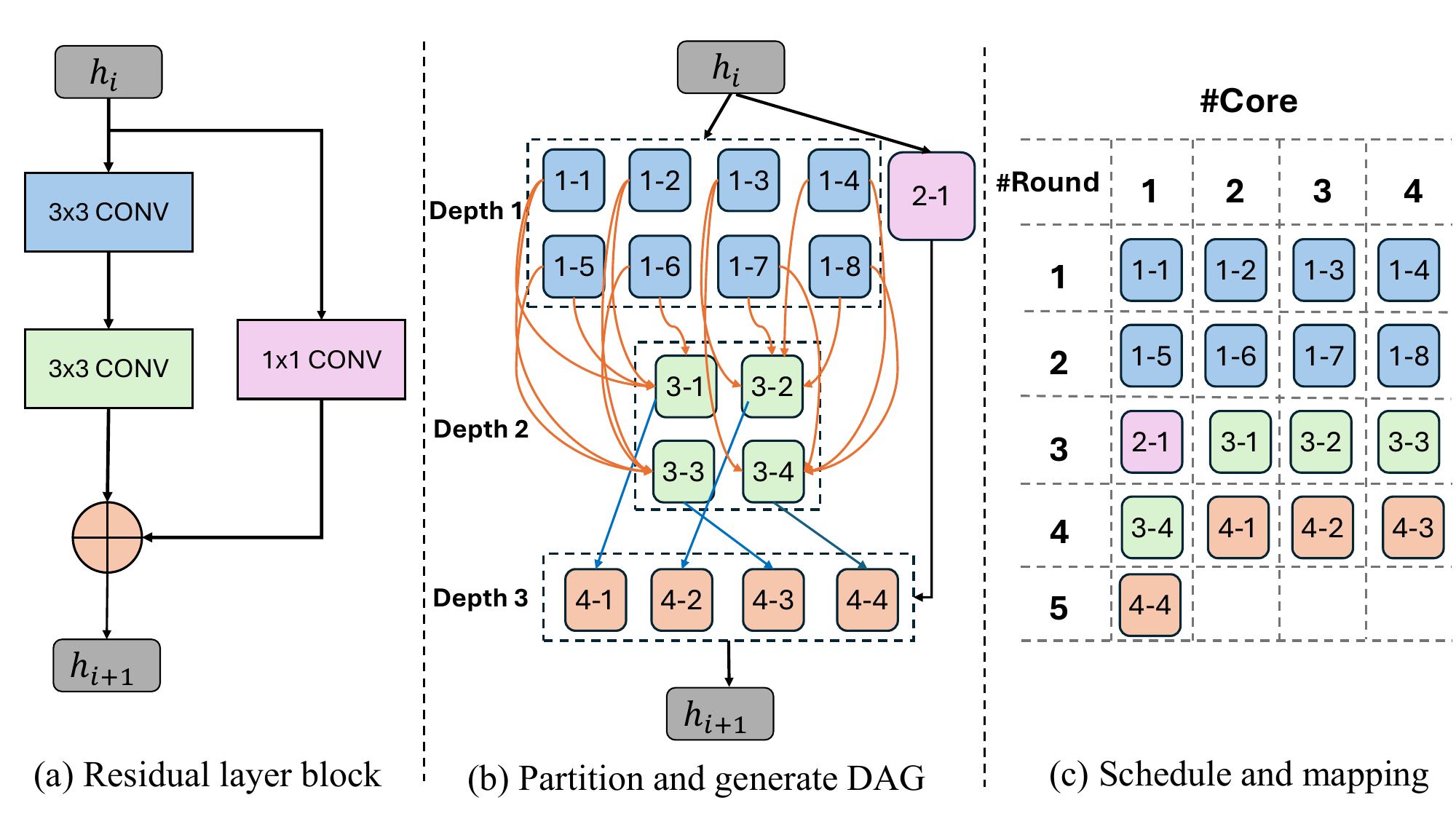}
\caption{The basic progress of task orchestration. The example is a ResNet block structure mapping into a tiled accelerator with $2\times2$ DNN cores.}
\label{fig_task_orch}
\end{figure}

The basic process, illustrated in Fig.~\ref{fig_task_orch}, includes four steps: (1) layer partitioning: partition the original layers into several $\mu$tasks. (2) $\mu$task-directed acyclic graph (DAG) generation. (3) scheduling: select a batch of $\mu$tasks at execution round. And (4) mapping: assign the selected $\mu$tasks to a specific NPU core.

In our implementation, we use utilization-directed partitioning and dynamic programming (DP)-based scheduling to reduce complexity. The original neural network (NN) workload is represented as a DAG, where each depth level contains one or more layers. During the partitioning stage, all layers at the same depth are initially divided into $n$ $\mu$tasks, where $n$ corresponds to the number of DNN cores. For each layer $l_i$, the number of assigned $\mu$tasks ($n_i$) is determined by its computational load ratio with respect to the total workload of that depth. When the $ubuf$ capacity is insufficient for the data volume of a $\mu$task, $n_i$ is increased to ensure feasibility.

For CONV and GEMM tensors, the partitioning priority for tensor dimensions follows $b \!\rightarrow\! h \!\rightarrow\! w \!\rightarrow\! c_o$ and $b \!\rightarrow\! h \!\rightarrow\! w$, respectively, where $b$, $h$, $w$, and $c_o$ denote batch size, height of OFM, width of OFM, and output channels. Hence, the $\mu$task DAG can be formulated as in Equations~\ref{eq_G}-\ref{eq_Edge}. 
\begin{gather}
G = \{V, E\}, \label{eq_G}\\
V = \{\mu task_{l,x,b}: [(h_s,h_e), (w_s,w_e), (c^i_s,c^i_e), (c^o_s,c^o_e)]\}, \label{eq_V}\\
E = \{e_{l,x,m,y}: \mu task_{l,x} \rightarrow \mu task_{m,y}\}, \label{eq_Edge} 
\end{gather}
where $l,\ m$ are the indices of layer in an NN, $x,\ y$ are the indice of the $\mu$tasks. In the DAG, each vertex $\mu task_{l,x,b}$ represents the $\mu$task $x$ in layer $l$ that processes the ouput feature map with the shape of height from $h_s$ to $h_e$, width from $w_s$ to $w_e$, input channel from $c^i_s$ to $c^i_e$, and output channel $c^o_s$ to $c^o_e$, respectively. while each edge encodes data dependencies among $\mu$tasks.

At each iteration, the scheduler selects at most $n$ $\mu$tasks to execute in parallel. The DP-based iterative search is adopted for $\mu$task scheduling, following~\cite{zheng2022atomic}, as it achieves high performance with low complexity. The scheduling priority is determined by the following rules:  
\textbf{(1)} $\mu$tasks from the same layer are prioritized to maximize reuse of input feature maps and weights already stored in on-chip buffers.  
\textbf{(2)} $\mu$tasks from layers at the same depth are scheduled earlier to release buffer capacity promptly, as they usually share common input data.  
\textbf{(3)} If some NPU cores remain idle after scheduling tasks of the current depth, additional $\mu$‑tasks with satisfied dependencies are then selected. This iterative process continues until all $\mu$tasks are executed.

Task mapping assigns each selected $\mu$task to a specific NPU core, which significantly impacts both performance and thermal characteristics. Prior studies~\cite{zheng2022atomic, bai2024klotski} employed dynamic programming or mixed-integer linear programming for mapping optimization, yet they neglected the thermal implications of the mapping decisions. Due to lateral heat conduction between cores, placing high-power tasks on centrally located cores can induce severe temperature hotspots.

To mitigate this effect, we introduce a greedy-based thermal-aware mapping strategy. NPU cores located near the center suffer the most from lateral heat accumulation, whereas corner cores experience the least. Consequently, the highest-power $\mu$tasks are preferentially mapped to the corner cores, while lower-power $\mu$tasks are assigned to the center cores.

This fine-grained task orchestration enables sophisticated data buffering mechanisms, including inter-core data reuse and data sharing. During each round, output data are written back to the uniform buffer for use in subsequent computations. Data sharing allows stored data to be reused by other cores when required. Compared with coarse-grained simulators such as TESA~\cite{shukla2023temperature} and Chiplet-Gym~\cite{mishty2024chiplet}, our fine-grained orchestration and buffering more accurately capture data-reuse opportunities, thereby substantially reducing DRAM accesses. A Last-in-Fist-out (LIFO) write-back policy is applied in our uniform buffer management: when insufficient space remains for incoming $\mu$tasks, the most recently stored output feature map data are iteratively written back to DRAM until sufficient buffer space is available.

\subsection{Performance, Power, Area, and Thermal Models}

\textbf{1) Latency Model:} 
The total delay $D$ is calculated as the sum of the delay of the target DNN workloads, as shown in Equation~\ref{eq_d_tot}. 
\begin{equation}
    D = \sum_{w=1}^{W} D_{w},  \label{eq_d_tot}
\end{equation}
where $W$ is the number of NNs in the workload and $D_w$ is the delay of an NN.
For each NN, given the generated DAG and task‑mapping scheme, its delay $D_w$ is obtained by summing the delays of all scheduling rounds, as shown in Equation~\ref{eq_d_nn}, 
\begin{equation}
        D_{w} = \sum_{r=1}^{R} D_r \label{eq_d_nn}, \\
\end{equation}
where $R$ denotes the total number of scheduling rounds and $D_r$ is the delay of a scheduling round.
The delay of each round is determined by the maximum latency among all NPU cores, as given in Equation~\ref{eq_D_r}, 
\begin{equation}
    D_r =\max\{D_1, D_2, \ldots, D_n\} \label{eq_D_r}, \\
\end{equation}
where $n$ is the number of NPU cores. The execution of an NPU core consists of three main phases: (1) fetching IFM and WEI data from DRAM or other NPU cores, (2) reading data to matrix unit or vector unit and writing back output into uniform buffer, and (3) writing the OFM data back to DRAM. 
A core can start computation only after all required data are loaded into local buffers. 
However, data write-back can be pipelined with computation. 
Therefore, the delay of an NPU core is given by Equation~\ref{eq_D_C},
\begin{equation}
    D_n = D_{in} + \max(D_{exe}, D_{out}) \label{eq_D_C},
\end{equation}
where $D_{in}$, $D_{exe}$, and $D_{out}$ represent the delays of input data movement, execution, and output write-back, respectively. 
Both $D_{in}$ and $D_{out}$ are related to data transfer bandwidth on DRAM and data volume of interconnect links.

For chiplet-based DNN accelerators, data transfers occur through two paths: the DRAM interface and the NoC/NoP interconnects. 
The overall data movement delay ($D_{move}$) is determined by the slower (maximum) phase between DRAM ($D_{DRAM}$) and NoC/NoP interconnects ($D_{nocp}$), as shown in Equation~\ref{eq_d_comm},
\begin{equation}
    D_{move} = \max\{D_{DRAM}, D_{nocp}\}.  \label{eq_d_comm}  
\end{equation}
The DRAM latency equals the total data volume divided by the bandwidth. For NoC/NoP, we moniter the data volume of each connection link. The latency of inter-core communication depends on the most congested NoC/NoP link, as shown in Equation~\ref{eq_d_nocp}, 
\begin{equation}
    D_{nocp} = \frac{Vol_{max}}{bw_{nocp}} \label{eq_d_nocp},
\end{equation}
where $Vol_{max}$ is the data volumn of most congested link and $bw_{nocp}$ is the bandwidth of NoC/NoP, consistent with prior works~\cite{zheng2022atomic, bai2024klotski, gao2019tangram, cai2023inter, cai2024gemini}.

\textbf{Detailed Core-Level Execution Model:} 
As shown in Fig.~\ref{fig_acc_arch}(b), the execution of an NPU core involves four main steps: 
(1) transferring input data from the uniform buffer to IFM and WEI buffers, 
(2) computation through matrix or vector units, 
(3) writing OFM data back to the uniform buffer, 
and (4) DRAM write-back. 

The data movement of different buffers is refered to SET~\cite{cai2023inter} and Gemini~\cite{cai2024gemini}. Due to the limited capacity of IFM and WEI buffers, only partial input data can be loaded at once. 
To maximize data reuse, the hardware selectively fixes either IFM or WEI data depending on reuse frequency and required buffer size. 
The maximal reuse factor is constrained by the depth of the accumulator register file (\textit{Acc RF}). 
Consequently, the corresponding buffer access volumes can be derived accordingly. 

Except for the uniform buffer, all on-chip buffers are double-buffered to enable overlapping data access and computation. 
Hence, the delay of NPU core ($D_{exe}$) execution comprises two components: computing delay $D_{comp}$ and on-chip data transfer delay $D_{buf}$. 
As presented in Equation~\ref{eq_d_core}, the effective core delay is determined by their maximum. 
\begin{equation}
    D_{core} = \max(D_{comp}, D_{ubuf}). \label{eq_d_core}
\end{equation}
The buffer delay equals the accessed data volume divided by the buffer port bit-width, as defined in Equation~\ref{eq_d_buf}. 
\begin{equation}
    D_{buf} = \frac{Vol}{bw_{buf}}, \label{eq_d_buf}
\end{equation}
where the $D_{buf}$ is the delay of on-chip buffer, $Vol$ is the data volume of access, and $bw_{buf}$ is the bit-wdith of this buffer.

Unlike previous works that neglect the utilization of matrix units, our model explicitly accounts for it. 
Each NPU core integrates a weight-stationary systolic-array matrix unit and a MAC-tree-based vector unit. 
The systolic array is configured as $h_{sa} \times w_{sa}$, while the vector unit comprises $w_{sa}$ computing units to maintain consistent bitwidth. 

For matrxi unit, the total computation delay is expressed in Equation~\ref{eq_d_comp}, 
\begin{equation}
    D_{comp} = \frac{\#ops}{\#PE \times u_{hsa} \times u_{wsa} \times u_{osa}}, \label{eq_d_comp} \\
\end{equation}
where the $\#ops$ is the total operations of GEMM and CONV computing, $\#PE$ is the number of PEs in systolic array, and $u_{hsa}\ ,u_{wsa}\ , u_{osa}$ are the utlizations of height, width, and output shape dimensions, respectively. In each computation batch, systolic arrays are constrained to process output matrix with shapes of $h_{sa} \times w_{sa}$. 
Non-aligned dimensions must be padded with zeros, following standard systolic array implementations~\cite{kung1979systolic, asgari2020meissa}. 
For CONV $\mu$tasks ($h\times w\times c_i\times c_o$) and GEMM tasks ($M\times K\times N$), utilization factors are defined in Equations~\ref{eq_u1_mtu}–\ref{eq_u3_mtu}. 
\begin{gather}
    u_{hsa} = 
    \begin{cases} \label{eq_u1_mtu}
        \frac{c_i}{h_{sa}} / \lceil \frac{c_i}{h_{sa}} \rceil, & \text{for CONV $\mu$tasks}\\
        \frac{K}{h_{sa}} / \lceil \frac{K}{h_{sa}} \rceil, & \text{for GEMM $\mu$tasks}
    \end{cases} \\
    u_{wsa} = 
    \begin{cases} \label{eq_u2_mtu}
        \frac{c_o}{w_{sa}} / \lceil \frac{c_o}{w_{sa}} \rceil, & \text{for CONV $\mu$tasks}\\
        \frac{N}{w_{sa}} / \lceil \frac{N}{w_{sa}} \rceil, & \text{for GEMM $\mu$tasks}
    \end{cases} \\
    u_{osa} = 
    \begin{cases} \label{eq_u3_mtu}
        \frac{h_ow_o}{h_{sa}} / \lceil \frac{h_ow_o}{h_{sa}} \rceil, & \text{for CONV $\mu$tasks}\\
        \frac{M}{h_{sa}} / \lceil \frac{M}{h_{sa}} \rceil. & \text{for GEMM $\mu$tasks}
    \end{cases}
\end{gather}
For weight-stationary systolic array, input channel of CONV $c_i$ and the column of left matrix $K$ of GEMM should align with the height of systolic array. The loop will execute $\lceil \frac{c_i}{h_{sa}} \rceil$ times. If the remaining input channel is less than the height of systolic array, part of PEs will not work. Similarly, the output channel of CONV $c_o$ and the column of right matrix $N$ should align with width of systolic array. In addition, $h_{o}\times w_{o}$ of CONV and row $M$ of GEMM should also align height of systolic array, since it can only handle $h_{sa}$ rows each time.

For element-wise and pooling layers executed on the vector unit, utilization follows the same formulation as Equation~\ref{eq_u2_mtu}, since there are $w_{sa}$ computing unit in vector unit.

\textbf{2) Energy Model:} 
The total energy consumption $E$ is the sum of energy consumption of target DNN workloads, as shown in Equation~\ref{eq_E_total}. 
\begin{equation}
    E = \sum_{w=1}^{W} E_{w}. \label{eq_E_total}
\end{equation}
For each NN, the energy cost $E_w$ is analytically modeled as the sum of DRAM, computation cores, and communication energy, as shown in Equation~\ref{eq_E}. 
\begin{equation}
    E_{w} = E_{DRAM} + E_{core} + E_{comm}. \label{eq_E} \\
\end{equation}
Communication energy includes both NoC and NoP components. 
Because NoC links are clock-gated when idle, their energy usage follows Equation~\ref{eq_E_module}. 
\begin{equation}
    E_{ops} = \#ops \times e_{ops} \label{eq_E_module}, \\ 
\end{equation}
where the $\#ops$ the number of a type of operation and $e_{ops}$ is energy cost of this type of operation.

For D2D links in NoP, energy consumption depends on the interconnect type. Clock-embedded D2D (e.g., SerDes~\cite{shivnaraine202111, shokrollahi201610, singh201426}) recovers the clock from data and consumes, hence the energy cost can be $E = \#D2D \times P_{D2D} \times \text{Latency}$. 
In contrast, clock-forwarding D2D (e.g., GRS~\cite{poulton20130, poulton20181}, UCIe~\cite{UCIe1.1}) employs a dedicated clock channel and can enter low-power states when idle; its energy is modeled similarly to on-chip NoC links using Equation~\ref{eq_E_module}. 
This work focuses on the latter type, consistent with existing systems such as Simba~\cite{shao2019simba}, NN-Baton~\cite{tan2021nn}, and Chiplet-Gym~\cite{mishty2024chiplet}. In addition to NoP, the energy cost of modules such as NoC, buffers, and computing unit, are also based on the Equation~\ref{eq_E_module}.
The average power is derived from the energy cost divided by the total latency, as shown in Equation~\ref{eq_P_avg}, where $m$ is a module in chiplet-based DNN accelerators, $D$ is the total delay, and $P_{clk}$ is the period of clock.
\begin{gather}
    P_{avg}^m = \frac{E_m}{D\times P_{clk}}. \label{eq_P_avg}
\end{gather}

\textbf{3) Area and Yield Models:} 
The total system area is computed as Equation~\ref{eq_A}, 
\begin{equation}
    A_{total} = 
    \sum_{i=1}^{xcut-1}\sum_{j=1}^{ycut-1} A_{ICS} 
    + \sum_{i=1}^{XX}\sum_{i=1}^{YY} A_{comp} 
    + A_{io}, \label{eq_A} 
\end{equation}
where $xcut$ and $ycut$ are the chiplet divisions along the X- and Y-directions, $XX$ and $YY$ are the core counts along these dimensions, and $A_{io}$ denotes the I/O die area. 
Each compute die area $A_{die}$ consists of the matrix unit ($A_{mtxu}$), vector unit ($A_{vecu}$), buffers ($A_{buf}$), D2D links ($A_{D2D}$), and other supporting modules ($A_{other}$), as formulated in Equation~\ref{eq_A_chip}. 
\begin{equation}
    A_{die} = A_{mtxu} + A_{vecu} + A_{buf} + A_{D2D} + A_{other}. \label{eq_A_chip} \\ 
\end{equation}
Following Chiplet-Gym~\cite{mishty2024chiplet}, we assume these auxiliary modules (control unit, NoC, router, DMA, etc.) account for approximately 30\% of the die area. 
The D2D link area is scaled proportionally to the link bandwidth based on GRS~\cite{poulton20181}. 
The I/O die area includes PHY modules such as PCIe, DDR, and D2D, whose sizes are derived from~\cite{cai2024gemini}. 

The die yield $Y$ follows the Negative Binomial model in Equation~\ref{eq_Y}, 
\begin{gather}
    Y = \left(1 + \frac{dA_{die}}{\alpha}\right)^{-\alpha}, \label{eq_Y}
\end{gather}
where $d$ denotes the defect density, $A_{die}$ the die area, and $\alpha$ the clustering parameter.

\begin{figure}[t]
    \centering
    \includegraphics[width=0.9\linewidth]{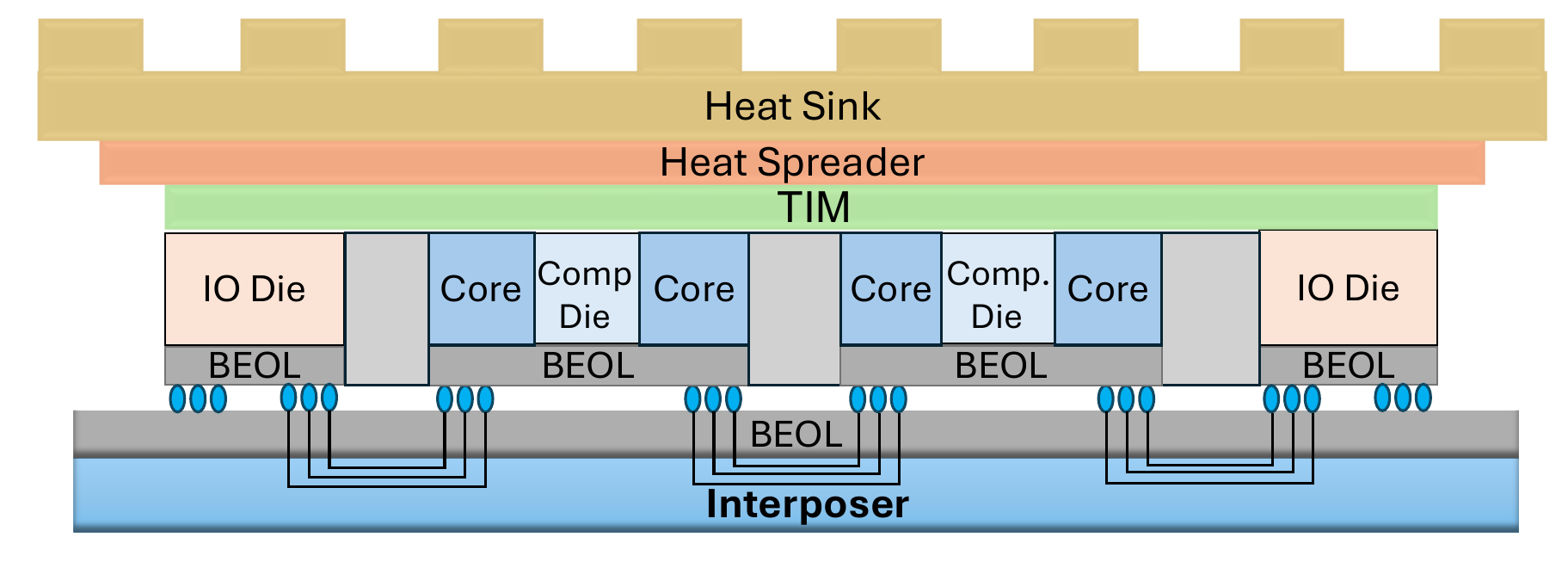}
    \vspace{-1mm}
    \caption{Cross-sectional view of a chiplet-based DNN accelerator. Each compute die contains multiple DNN cores connected via an on-die NoC, while die-to-die links are realized through micro-bumps, BEOL interconnects, and an interposer.}
    \label{fig_thermal_view}
    \vspace{-2mm}
\end{figure}

\textbf{4) Thermal Model:} 
Thermal behavior is simulated using HotSpot~6.0~\cite{zhang2015hotspot}, taking as input the component-wise average power, floorplan, and material properties. 
Fig.~\ref{fig_thermal_view} shows a cross-sectional view of the chiplet-based DNN accelerator. 
Within each compute die, multiple NPU cores communicate through a mesh NoC. 
The die-to-die interconnects (NoP) are built in the interposer. 
Below the compute layers, thermal interface material (TIM) and a heat spreader conduct heat to the external heat sink. 
We assume NoP power is spatially distributed across the interposer. 
The thermal properties of TIM, heat spreader, and heat sink are adopted from~\cite{shukla2023temperature} for realistic modeling.

\begin{table}[t]
    \caption{Design space of scalable chiplet-based DNN accelerator.}
    \centering
    \begin{tabular}{c|c|c}
    \toprule
       \textbf{Variable} & \textbf{Parameter} & \textbf{Value Range} \\ \midrule
        $XX$ & DNN cores in X dimension  & 1–8 (step 1) \\
        $YY$ & DNN cores in Y dimension & 1–8 (step 1) \\
        $Xcut$ & Chiplet partition in X dimension  & 1–8 (step 1) \\
        $Ycut$ & Chiplet partition in Y dimension  & 1–8 (step 1) \\
        $h_{sa}$ & Systolic array height  & 16–256 (step 16) \\
        $w_{sa}$ & Systolic array width  & 16–256 (step 16) \\
        $ICS$ & Inter-chiplet spacing & 0.5–3.5\,mm (step 0.3\,mm) \\
        $ubuf$ & uniform buffer size & 128–8192\,KB (power of 2) \\
        $NoCP_{bw}$ & NoC/NoP bandwidth & 16–256 (step 16) \\ 
    \bottomrule
    \end{tabular}
    \label{tab:design_space}
    \vspace{-2mm}
\end{table}

\subsection{Early-Stage Design Space Exploration}
In practical chiplet-based system design, the interposer area is typically limited to control manufacturing cost. Meanwhile, peak temperature constraints must be evaluated at an early stage to ensure thermal reliability. Hence, our proposed DSE framework adopts the maximum allowable interposer area and peak temperature as system constraints. Each design configuration must satisfy both constraints simultaneously.

Designing a chiplet-based DNN accelerator involves multiple trade-offs. Intuitively, increasing the number of computing cores can improve throughput, yet excessive expansion may reduce array utilization due to zero padding once the systolic array height and width exceed optimal limits. Conversely, partitioning a large NPU core into smaller chips improves utilization and yield, but also increases inter-core and inter-chip communication overhead. At the chiplet level, increasing the inter-chiplet spacing alleviates lateral heat coupling but occupies more area. Therefore, an efficient DSE algorithm is essential for identifying optimal architectural configurations that balance these trade-offs.

The DES propblem is formulated in Equations~\ref{eq_cost}-\ref{eq_dse}.
\begin{gather}
    EDYP = E \times D \times Y^{-1}, \label{eq_cost} \\
    \underset{x\in \Omega}{\arg\min} \ EDYP(x)
    \quad \text{s.t.} \quad
    c_1(x) \leq C_1,\ 
    c_2(x) \leq C_2, \label{eq_dse}
\end{gather}
where $D$, $E$, $Y$, $C_1$, and $C_2$ denote the total delay, total energy consumption, die yield, peak-temperature constraint, and interposer area constraint, respectively. The DSE engine seeks the optimal design point $x$ within the design space $\Omega$ that minimizes $EDYP$ under the given physical constraints. In mobile and edge scenarios, a lower energy–delay product indicates higher energy efficiency and extended battery life. Accordingly, our DSE objective is formulated as the product of delay, energy, and yield inversely, defined in Equation~\ref{eq_cost}.

\textbf{Problem (Chiplet-Based DNN Accelerator Design Exploration).} 
\textit{Given a design space $\Omega$, each DNN accelerator architecture is represented as a feature vector $x$. The delay–energy–yield triplet ($D,E,Y$) forms the $EDYP$ objective space $\mathcal{Y}$, while peak temperature $c_1$ and interposer area $c_2$ constitute the constraint space $\mathcal{C}$. Through our simulator, for any $x$, the values of $y \in \mathcal{Y}$ and $\{c_1, c_2\}\in \mathcal{C}$ are derived. The optimization goal is to find a set of feature vectors $\mathbf{X}$ minimizing $EDYP$ while all $\mathbf{x}\in\mathbf{X}$ satisfy $c_1(\mathbf{x})\le C_1$ and $c_2(\mathbf{x})\le C_2$.}

Table~\ref{tab:design_space} summarizes the configurable parameters in our DSE framework, including both chiplet-level and core-level architectural variables. To reduce dimensionality, parameters with low importance are fixed. In detail, the number of vector units equals $ w_{sa}$. The capacity of WEI, IFM, and OFM buffers is set to one-fourth of the uniform buffer, following prior designs such as Simba~\cite{shao2019simba} and DaVinci~\cite{liao2019davinci}. The entire design space contains up to \textit{one billion} potential points. Moreover, incorporating thermal and area constraints increases the complexity of searching for the optimal configuration. To efficiently explore this vast space, we propose \textbf{ThermoDSE}, an agile and effective optimization framework composed of two phases: (1) active-learning-based initialization and (2) scalable constrained Bayesian optimization.

\begin{algorithm}[t]
\caption{\textbf{TED}($\mathbf{U}, \mu, b$)}
\label{alg_ted}
\begin{algorithmic}[1]
\STATE \textbf{Input:} Unsampled design set $\mathbf{U}$, normalization coefficient $\mu$, and sample count $b$
\STATE \textbf{Output:} Representative sampled subset $\chi$ with $|\chi|=b$
\STATE Initialize $\chi \leftarrow \emptyset$, compute kernel matrix $K_{\mathrm{uu'}}=f(u,u')$, $\forall u,u'\in \mathbf{U}$
\FOR{$i=1$ to $b$}
    \STATE Select $x_{*}=\underset{u\in\mathbf{U}}{\arg\max}\,\text{Tr}[K_{\mathbf{U}x}(K_{xx}+\mu\mathbf{I})^{-1}K_{x\mathbf{U}}]$
    \STATE $\chi \leftarrow \chi \cup x_{*}$, $\mathbf{U}\leftarrow\mathbf{U}\setminus x_{*}$
    \STATE $K \leftarrow K -K_{\mathbf{U}x_*}(K_{x_*x_*}+\mu\mathbf{I})^{-1}K_{x_*\mathbf{U}}$
\ENDFOR
\RETURN Sampled initialization set $\chi$
\end{algorithmic}
\end{algorithm}
\subsubsection{\textbf{Active Learning for Initialization}}
For Bayesian Optimization, the initial data set can significantly affect the search quality and efficiency. Therefore, obtaining a well-distributed initial dataset is critical for accelerating convergence in subsequent optimization. Two principles guide the initialization process: (1) the sampled feature vectors should uniformly cover the design space, and (2) their diversity should adequately represent the architectural variability of that space.


Recently, active learning has been widely adopted in DSE due to its ability to generate high-quality and information-rich initial datasets efficiently. The transductive experimental design method~\cite{liu2013learning, wu2019active} has shown promising results in various domains, including high-level synthesis~\cite{pouget2025automatic}, CPU microarchitecture tuning~\cite{fan2024explainable}, and LLM accelerator design~\cite{liu2025llmshare}. Inspired by these, we apply TED for the initialization of chiplet-based DNN accelerator DSE.

TED selects representative design samples that maximize information coverage by diversifying their mutual distances. Given an unsampled pool $\mathbf{U}$, the algorithm iteratively selects a subset $\chi$ from $\mathbf{U}$ that maximizes intra-set diversity. As shown in LINE 4-7 of Algorithm~\ref{alg_ted}, TED will select a feature vector $x_*$ that can maximize the diversity of sampling set from the unsample space at each iteration.


\subsubsection{\textbf{Scalable Constrained Bayesian Optimization}}
The DSE algorithm must satisfy two key requirements: (1) efficiently locate the optimal design within a limited number of iterations, and (2) explore feasible solutions near constraint boundaries.
The SCBO framework~\cite{eriksson2021scalable} fulfills both objectives by focusing on critical areas through transformations of the objective and constraints. We thus adopt SCBO as the optimization engine for ThermoDSE.

\begin{algorithm}[t]
\caption{\textbf{SCBO($\Omega,C_1,C_2,l,n$)}}
\label{alg:scbo}
\begin{algorithmic}[1]
\STATE \textbf{Input:} Design space $\Omega$, objective function $f(\mathbf{x})$, constraint functions $\{c_i(\mathbf{x})\}_{i=1}^{2}$, iteration count $T$, batch size $n$, constraints $C_1$, $C_2$
\STATE \textbf{Initialization:} $\mathbf{x}\leftarrow TED(\Omega,\mu,b)$
\STATE Evaluate $EDYP$ and $\{c_i(\mathbf{x})\}$
\FOR{$t=1$ to $T$}
    \STATE Train GP surrogate models for $f(\mathbf{x})$ and $\{c_i(\mathbf{x})\}$
    \STATE Update trust-region length $l$ via Equation~\ref{eq_ltr}
    \STATE $x_c = \arg\min_{\mathbf{x}} f(\mathbf{x})\ \text{s.t.}\ c_1(\mathbf{x})\le C_1,\ c_2(\mathbf{x})\le C_2$
    \STATE $\mathbf{x}_{next} \leftarrow CMPS(GP_{EDYP}, \{GP_{c_i}\}, x_c, l, n)$
    \STATE Evaluate $EDYP$ and constraints for $\mathbf{x}_{next}$
\ENDFOR
\RETURN Best feasible configuration $\mathbf{x}^{*}$ satisfying $c_i(\mathbf{x}^{*})\!\le\!C_i$
\end{algorithmic}
\end{algorithm}

Algorithm~\ref{alg:scbo} outlines the SCBO workflow. Gaussian Process (GP) surrogate models are employed to fit the objective (EDYP and the constraints) based on evaluated design points $\{(x_i, f_i)|i=1, \dots, n\}$ and predict the future design points $\{(x^*_i, {f^*}_i)|i=1, \dots, m\}$. GP models offer a non-parametric regression capability and perform well with sparse samples. The GP assumes a multivariate Gaussian distribution as the prior distribution on the objective function $f(\mathbf{x})\sim\mathcal{GP}(\mu, \Sigma)$ with mean $\mu = m(\mathbf{x})$ and covariance matrix $\Sigma$ calculated via the kernel function $K$. Note that we employ a scaled Matérn kernel for both objectives and constraints to balance smoothness and adaptability. Given evaluated design points $\mathbf{X}$ and the design points we would like to test next $\mathbf{X^*}$, we form the joint distribution as shown in Equation~\ref{eq_gp}. 
\begin{align}
    \left[\stackanchor{\mathbf{f}} {\mathbf{f^*}} \right] \sim 
    \mathcal{N} \left( \stackanchor
    {m(\mathbf{X})}{m(\mathbf{X^*})}, 
    \left[ \stackanchor{K(\mathbf{X,X}), \ K(\mathbf{X, X^*}) }
    { K(\mathbf{X^*, X}), \ K(\mathbf{X^*, X^*})} \right] \right).
    \label{eq_gp}
\end{align}


Then we can obtain the posterior distribution of $f^*$ conditioning on known information, as shown in Equation \ref{eq_gp_pos}. The posterior distribution provides us with insight into the potential of future samples, helping us to make more reasonable decisions.
\begin{equation}
\begin{split}
\label{eq_gp_pos}
    \mathbf{f^*} &| \mathbf{X^*}, \mathbf{X}, \mathbf{f} \sim \mathcal{N}(\mu_{f^{*}}, \Sigma_{f^{*}}), \text{where}\\
    &\mu_{f^{*}} = m(\mathbf{X^*}) +K(\mathbf{X^*, X})K(\mathbf{X,X})^{-1} (\mathbf{f} - m(\mathbf{X})),\\
    &\Sigma_{f^{*}} = K(\mathbf{X^*, X^*}) - K(\mathbf{X^*, X})K(\mathbf{X,X})K(\mathbf{X, X^*}).
\end{split}
\end{equation}

Conventional BO methods rely on acquisition functions such as Expected Improvement or Upper Confidence Bound to evaluate the potential of future samples over the posterior distributions. To be more detailed, they randomly sample several starting points and perform line search to find the sample that maximizes the acquisition function. In vast high-dimensional spaces, the line search is as difficult as looking for a needle in a haystack, resulting in low efficiency.

To improve the efficiency of high-dimensional optimization, SCBO defines the trust regions and performs separate local optimization runs simultaneously to avoid getting lost in the overly large design space. The trust regions are hyper-rectangles centered at the current best design. In each iteration, the trust-region length $l$ is adjusted adaptively according to Equation~\ref{eq_ltr} based on consecutive successes or failures, where $l_{max}$ and $l_{min}$ denote predefined bounds.

\begin{gather}
l =
\begin{cases}
\min(2l,\,l_{max}), & \text{if } c_{success} > th \\
\max(l/2,\,l_{min}), & \text{if } c_{fail} > th
\end{cases}
\label{eq_ltr}
\end{gather}

A success is registered when the objective improves beyond a threshold while satisfying constraints; otherwise, a failure is recorded. When the success/fail counter reaches the user-defined threshold ($th$), the trust-region length $l$ will be adjusted. The current best feasible design then becomes the new center $\mathbf{x}_c$ for local exploration. The candidate sampling space is defined in Equation~\ref{eq_tr}, where $\epsilon$ adds a stochastic perturbation to encourage exploration and prevent local optimal trapping.

\begin{gather}
\mathcal{X}_{\text{tr}}(\mathbf{x}_c, l) =
\bigl\{
\mathbf{x}\in\Omega \mid \|\mathbf{x}-\mathbf{x}_c\| \le l+\epsilon
\bigr\}.
\label{eq_tr}
\end{gather}

SCBO uses constrained maximum posterior sampling (CMPS) to generate the next candidate batch from the dynamic sampling region. For each sampled realization of objective function, peak-temperature constraint, and area constraint($\hat{f}(x_i),\hat{c_1}(x_i),\hat{c_2}(x_i)$), if feasible points exist within the trust region (i.e., $\hat{F}=\{x_i|\hat{c_1}(x_i)\!<\!C_1,\ \hat{c_2}(x_i)\!<\!C_2\}$), SCBO selects $x^*=\arg\min_{x\in\hat{F}}\hat{f}(x)$. Otherwise, it selects the configuration with the minimum total constraint violation.

\section{Experiment}
\subsection{Experiment Setup}
We conduct comprehensive experiments to evaluate the proposed framework on an Intel(R) Xeon(R) Gold 6246R CPU @ 3.40GHz with 128GB of memory. For hardware configurations, the energy and area of the D2D interconnect are set to $1.17\ pJ/bit$ and $81K\ \mu m^2$ per 25Gb/s bandwidth, respectively, as reported in \cite{poulton20181}. Following previous work~\cite{cai2024gemini}, we adopt a 4-channel DRAM as the off-chip memory, providing a total capacity of 4GB and a peak bandwidth of 128GB/s. The energy consumption of the NoC and DRAM is set to $0.6\ pJ/bit$ and $8\ pJ/bit$, respectively~\cite{zheng2022atomic, cai2024gemini}. We implement an INT-8 MAC module and an FP-16 MAC module in Verilog HDL as the fundamental processing elements of the matrix unit and vector unit, respectively. The synthesized energy and area data are obtained using Synopsys Design Compiler with the TSMC 28nm process. Similarly, buffer configurations are derived from the TSMC memory compiler.  

To ensure a fair comparison with baselines, we assume a 14nm process node and scale the results from the 28nm process to 14nm using DeepScaleTool~\cite{deepscaletool}. In the yield model, the parameters $\alpha$ and $d$ are set to 10 and 0.08, respectively, following Chiplet-actuary~\cite{feng2022chiplet}. In addition, we assume an operating frequency of 1.8GHz (the peak frequency used in Simba~\cite{shao2019simba}), since early-stage thermal exploration should consider the worst-case thermal conditions.

This work focuses on embedded devices, where cooling capacity is typically limited. Accordingly, the interposer area is constrained to $300\,mm^2$, and the maximum peak temperature is limited to \SI{75}{\degreeCelsius}. For thermal simulation, we use the default HotSpot~6.0 ambient temperature of \SI{45}{\degreeCelsius}, with a convection resistance of 0.4 K/W to represent restricted cooling conditions~\cite{skadron2003temperature}.

\subsection{Simlation Validation}

\begin{table}[t]
    \centering
    \caption{Benchmark for COMMON NNs in AR/VR.}
    \begin{tabularx}{1.05\linewidth}{c|Y|Y|Y|Y|Y|Y}
    \toprule
     & \textbf{ResNet-50} & \textbf{Goog-LeNet} & \textbf{Mobile-Netv2} & \textbf{Yolo-v2} & \textbf{U-Net} & \textbf{BERT}\\ 
    \midrule
     Batch size     & 2    & 2   & 4   & 4   & 2  & 1  \\
     FM Vol.~(MB)   & 15.6 & 3.5 & 6.5 & 5.0 & 27.6&  52.3 \\
     Wei. Vol.~(MB) & 24.3 & 6.6 & 1.7 & 18.9 & 13.4&  32.5\\
     GOPS        & 3.6  & 1.5 & 0.24& 2.6 & 20.8 & 15.8 \\ 
     Sparsity       & 0.22 & 0.38 & 0.37 & 0.01 & 0.45 & 0 \\
     \bottomrule
    \end{tabularx}
    \label{tab:benchmark}
    \vspace{1mm}
    \parbox{\linewidth}{
        \textbf{Note:} The input image size for all networks except BERT is $224\times224\times3$. BERT refers to the BERT-Small model.
    }
    \vspace{-3mm}
\end{table}
\subsection{Implementation Details}
\textbf{Architecture/Model:} We evaluate multiple architectures and corresponding DSE algorithms in our experiments. Four representative architectures are considered:  
1) \textbf{Simba-like:} This design follows Simba~\cite{shao2019simba} and is implemented within our framework. It consists of $6\times6$ chiplets connected through a NoC/NoP network, with fine-grained task orchestration enabling data reuse and sharing.  
2) \textbf{Arch-T:} Derived from TESA~\cite{shukla2023temperature}, this design has no NoC/NoP interconnect between NPUs or chiplets, with each chiplet operating independently. We implement it by disabling NoC/NoP interconnects in our simulator.  
3) \textbf{Arch-G:} Based on the analytical model used in Chiplet-Gym~\cite{mishty2024chiplet}. Although the NoP provides inter-chiplet connectivity, no data reuse or sharing is supported. This configuration is represented in our framework by disabling data buffering.  
4) \textbf{Arch-S:} A chiplet-based DNN accelerator architecture supporting fine-grained task orchestration across multiple chiplets.

\textbf{DSE Algorithm:} Three DSE algorithms are implemented:  
1) \textbf{SA} is adopted from TESA~\cite{shukla2023temperature} to search for feasible chiplet-based DNN accelerator designs under thermal and area constraints.  
2) \textbf{RL} is implemented following Chiplet-Gym~\cite{mishty2024chiplet}, using the Gymnasium framework~\cite{towers2024gymnasium}. Proximal Policy Optimization (PPO) employs Multi-Layer Perceptrons (MLPs) for both the policy and value networks. The policy (actor) network has the architecture [10, 64, 64, 810], and the value (critic) network is [10, 64, 64, 1], both with \texttt{tanh} activation functions. Since the original RL algorithm does not consider design constraints, we refine the reward function to incorporate area and temperature penalties, ensuring feasible final designs.  
3) \textbf{TED+SCBO:} Our DSE algorithm, implemented using BoTorch~\cite{balandat2020botorch}.

\textbf{Benchmarks:} Table~\ref{tab:benchmark} is the summary of our benchmark, including the batch size, data volume of feature map, data volume of weight, and sparsity. The sparsity refers to previous work~\cite{pao2025eda} and can significantly influent the computing energy cost. These evaluation targets AR/VR applications~\cite{kwon2021heterogeneous}, covering a diverse set of networks: image segmentation (U-net, Yolov2), object recognition (ResNet50), object detection (MobileNetv2, GoogLeNet), and speech recognition (BERT). The total delay and energy are computed as the weighted sum of each network’s corresponding batch execution.

\subsection{Overall Comparison}
We comprehensively compare \textbf{ThermoDSE} with prior works, including Simba~\cite{shao2019simba}, TESA~\cite{shukla2023temperature}, and Chiplet-Gym~\cite{mishty2024chiplet}. The experiments include five configurations: Simba-like, TESA, TESA$^*$, Chiplet-Gym, and ThermoDSE.  
\textbf{Simba-like (B1)} mirrors the structure of Simba with $6\times6$ chiplets. Each chiplet integrates a $64\times16$ systolic array as the matrix unit and a 512KB uniform buffer.  
\textbf{TESA (B2)} and \textbf{TESA$^*$ (B3)} correspond to Arch-T with the SA algorithm. Due to the absence of NoC/NoP, Arch-T only supports network-level parallelism. 
\textbf{Chiplet-Gym (B4)} combines Arch-G with the RL-based DSE algorithm.

Two practical scenarios are considered: \textbf{TESA} represents the \emph{edge scenario}, while \textbf{TESA$^*$} denotes an \emph{ideal scenario}.    
In the edge scenario, workloads are limited; hence, we fix the number of chiplets in \textbf{TESA} to six—matching the number of network types—to avoid over-provisioning, following~\cite{shukla2023temperature}. During execution, each network is mapped to a dedicated chiplet, and the scheduler evenly distributes workloads across chiplets to prevent thermal hot spots. The overall delay is determined by the maximum latency among all networks.  
In contrast, \textbf{TESA$^*$} assumes ample workloads for full parallel execution, with the total delay estimated by dividing the sequential delay by the number of NPU cores.  

\begin{table*}[t]
    \centering
    \caption{Overall comparison of peak temperature and performance for different architectures and optimizers}
    \begin{tabular}{c|>{\centering\arraybackslash}p{8cm}|c|c|c|c|c}
    \toprule
    \textbf{} & \textbf{Architecture Description} & \textbf{Temp.(\textcelsius)} & \textbf{Energy(mJ)} & \textbf{Delay(ms)} & \textbf{Yield} & \textbf{EDYP cost} \\ \midrule
     B1 & i) $6\times6$ NPU cores, $6\times6$ chiplets, 1.7mm ICS, 100GB/s NoP bandwidth. ii) 512KB uniform buffer and $64\times16$ systolic array. & 69.6 & 103.3 & 4.46 & 0.998 & 461.6 \\ \midrule
    B2 &  i) $3\times2$ NPU cores, $3\times2$ chiplets, 0.5mm ICS. ii) 4096KB uniform buffer and $160\times208$ systolic array. & 58.8 & 77 & 4.72 & 0.985 &  228.4\\ \midrule
    B3 &  i) $8\times2$ NPU cores, $8\times2$ chiplets, 0.5mm ICS. ii) 4096KB uniform buffer and $128\times128$ systolic array. & 71.1 & 63.2 & 2.46 & 0.989 &  157.2\\ \midrule
    B4 & i) $4\times4$ NPU cores, $4\times4$ chiplets, 1.1mm ICS, 96GB/s NoP bandwidth. ii) 1024KB uniform buffer and $160\times160$ systolic array. & 63.8 (74.6) & 107.6 (70.2) & 4.07 (2.35) & 0.991 & 441.9 (166.5) \\ \midrule
    \textbf{Ours} & i) $5\times4$ NPU cores, $1\times2$ chiplets, 2.3mm ICS, 240GB/s NoC bandwidth. ii) 1024KB uniform buffer and $208\times112$ systolic array. & 74.8 & 59.8 & \textbf{2.01} & 0.911 & \textbf{132.2} \\ \bottomrule
    \end{tabular}
    \vspace{3pt} \par
    \textbf{Note:} For Chiplet-Gym (B4), the values in parentheses are the results obtained with fine-grained modeling. 
    \label{tab:overall_comparison}
    \vspace{-2mm}
\end{table*}

Table~\ref{tab:overall_comparison} summarizes the optimal designs obtained from each optimizer under the constraints of \SI{75}{\degreeCelsius} peak temperature and $300\, mm^2$ area. The design produced by our framework consists of $5\times4$ NPU cores partitioned into 2 chiplets connected via a 240GB/s inter-chiplet link. The ICS is set to 2.3mm for improved cooling. Each NPU core employs a 1024KB uniform buffer and a $208\times112$ systolic array as the matrix computation unit. Owing to the intelligent resource allocation from our DSE algorithm, the achieved theoretical peak throughput is $1.8\text{–}12.5\times$ higher than the compared baselines.

\smallskip
\noindent
\textbf{1) Comparison with Simba-like (B1):}  
ThermoDSE achieves reductions of $1.7\times$, $2.2\times$, and $3.4\times$ in energy, delay, and EDYP cost, respectively. Because D2D interconnects are significantly more energy-consuming than NoC links, ThermoDSE tends to reduce the number of chiplets while increasing per-chiplet compute density to lower delay, even at a minor yield cost.

\smallskip
\noindent
\textbf{2) Comparison with TESA (B2) and TESA$^*$(B3):}  
Although Arch-T avoids data communication energy by design, ThermoDSE still achieves $1.7\times$ and $1.2\times$ lower EDYP cost than B2 and B3, respectively. In Arch-T, the absence of NoC/NoP means data reuse is restricted by the capacity of the local uniform buffer. As a result, the DSE algorithm allocates larger buffers to increase reuse. However, larger SRAMs introduce more energy consumption during access~\cite{muralimanohar2009cacti}. Interestingly, the energy increase in TESA due to larger SRAM closely matches the communication energy in our ThermoDSE design.  
Even in the ideal scenario (B3), where 100\% core utilization is achieved, its delay remains higher than that of ThermoDSE due to the higher computing and DRAM access latency. Moreover, it is impractical for each chiplet to have its own independent DRAM channel in a large chiplet system.

\smallskip
\noindent
\textbf{3) Comparison with Chiplet-Gym (B4):}  
Results indicate that the analytical model used in Chiplet-Gym is highly inaccurate for chiplet-based DNN accelerators. Fine-grained task orchestration can substantially improve both energy and delay. The errors in temperature, energy, and delay reach 9.8\textcelsius, 53\%, and 73\%, respectively. Consequently, the resulting design deviates considerably from the true optimal solution. Furthermore, due to inaccurate early-stage performance estimation, peak temperature is underestimated, leading to insufficient cooling design and misjudged performance margins.

\begin{figure*}[t]
    \centering
    \includegraphics[width=0.95\linewidth]{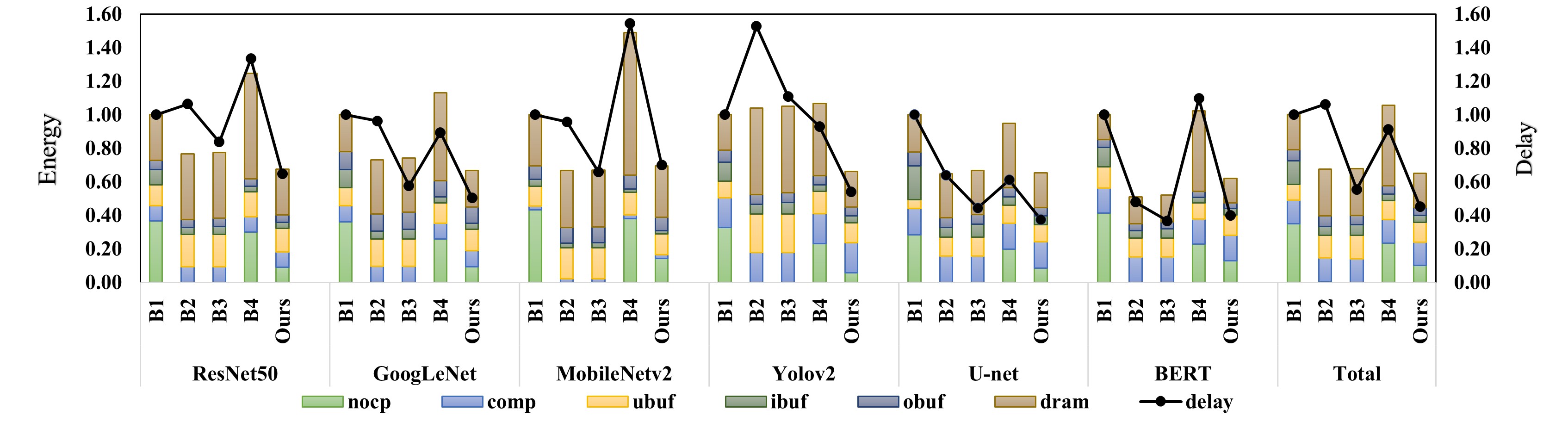}
    \caption{Normalized energy and delay across different architectures and optimizers (normalized to the Simba-like baseline).}
    \label{fig_compare_detail}
    \vspace{-4mm}
\end{figure*}
\begin{figure}[t]
    \centering
    \includegraphics[width=0.95\linewidth]{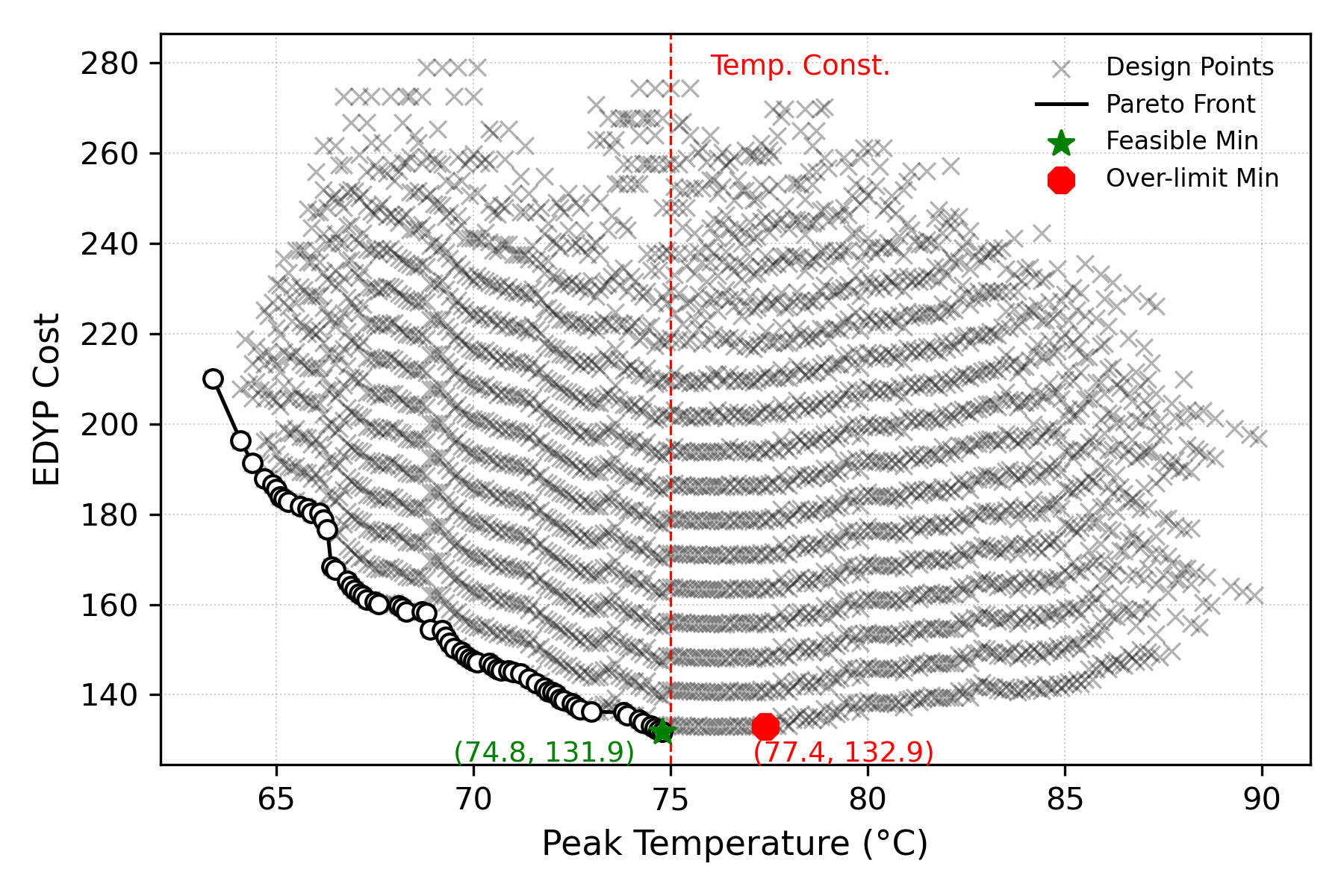}
    \vspace{-4mm}
    \caption{EDYP cost vs. peak temperature. Selected design points for chiplet-based DNN accelerators under $300\,mm^2$ area and 75\,\textcelsius{} thermal constraints.}
    \label{fig_scan_design}
    \vspace{-2mm}
\end{figure}
Fig.~\ref{fig_compare_detail} presents the normalized energy and delay results for each neural network. Chiplet-Gym (B4) exhibits the highest energy consumption, as inputs and outputs must be read and written to DRAM. Although its NoP-related energy is lower due to fewer chiplets than Simba-like design, the overall DRAM cost dominates. In contrast, ThermoDSE reduces both energy and peak temperature by using only two chiplets and minimizing expensive D2D interconnections.

The layer type of different NNs will mainly affect the performance of chiplet-based DNN accelerator. the workloads can be categorized into two groups:  
(1) \textbf{Compute-intensive networks}, such as ResNet50, GoogLeNet, and Yolov2, contain predominantly compute-heavy layers, creating strong opportunities for data reuse and sharing. Even though thermal-aware mapping introduces additional data movement, ThermoDSE still achieves lower delay and energy than B3. DRAM costs in B2 and B3 remain much higher than in ThermoDSE and Simba-like, as repeated weight transfers among independent chiplets incur significant DRAM energy overhead.  
(2) \textbf{Data-intensive networks}, such as MobileNetv2, U-Net, and BERT, contain layers dominated by data movement. For instance, depthwise convolutions in MobileNet offer limited data reuse, leading to slightly higher energy and delay in ThermoDSE compared with B3. Similarly, U-Net’s upsampling and BERT’s GEMM layers exhibit lower data reuse and higher data movement. Although the NoC/NoP interconnect provides larger buffers for intermediate reuse, the additional interconnect energy partially offsets DRAM savings.

For BERT, the delay of ThermoDSE is roughly 10\% higher than B3 with ideal scenario . This is primarily because smaller matrix multiplications lower the processing element utilization within the matrix unit. Moreover, BERT’s multi-head attention mechanism introduces extensive data traffic: the Q, K, and V computations are distributed across different NPU cores, and subsequent QK, QKV, and concatenation operations require intensive inter-core data fusion. Consequently, the NoC/NoP incurs higher latency. Adaptive, data-aware task mapping may further mitigate this overhead.  
Nevertheless, since BERT represents only one among several AR/VR workloads, ThermoDSE still achieves an overall $1.2\times$ lower delay than B3.
\subsection{DSE Algorithm Evaluation}
\begin{figure}[t]
    \centering
    \includegraphics[width=0.95\linewidth]{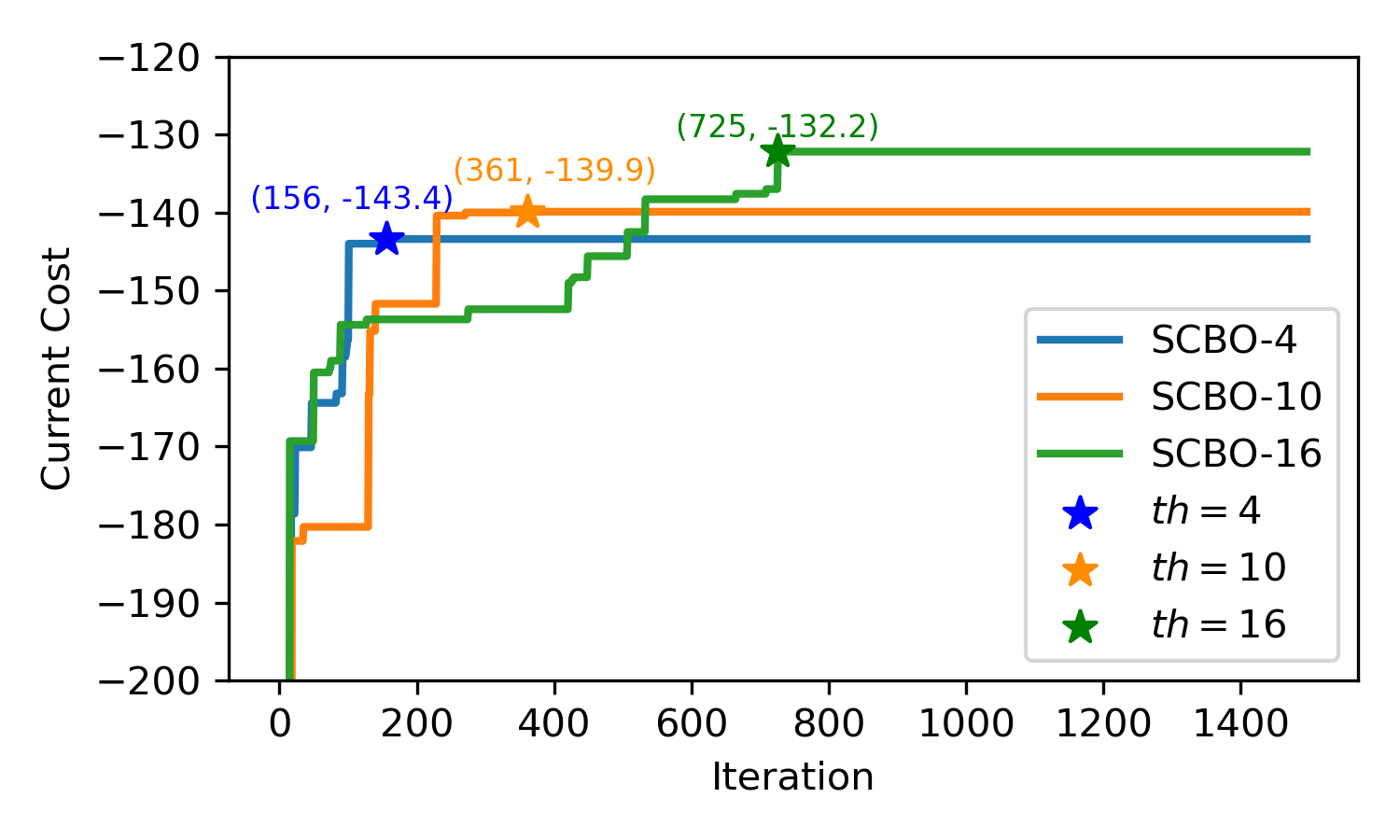}
    \vspace{-3mm}
    \caption{Convergence behaviors for different thresholds in SCBO.}
    \label{fig_scbo_cove}
    \vspace{-2mm}
\end{figure}
To better understand the relationship between thermal characteristics and EDYP, we conduct a exhaustively scan of approximately 440K neighboring design points around the optimal configuration obtained from ThermoDSE to identify the global optimal point. Fig.~\ref{fig_scan_design} illustrates the selected design points of chiplet-based DNN accelerators from this search.  
The best feasible design corresponds to a $5\times4$ NPU core array partitioned into $1\times2$ chiplets, connected via 256GB/s inter-chiplet links with an ICS of 2.3\,mm. Each core integrates a $208\times112$ matrix unit and a 1024KB uniform buffer. This design is nearly identical to the final configuration obtained by our DSE method, except that it adopts a higher interconnect bandwidth. The 256GB/s bandwidth slightly reduces both delay and overall EDYP cost compared with the final DSE-derived design.  

In addition, the curve in Fig.~\ref{fig_scan_design} exhibits a U-shaped trend. Initially, as hardware resources increase, the EDYP cost decreases due to shorter delays. However, beyond a certain point, over-provisioning NPU cores leads to a substantial rise in data communication energy, while delay improvement becomes negligible, resulting in higher EDYP cost. Meanwhile, the $300\,mm^2$ interposer area provides sufficient thermal dissipation capacity, ensuring that the best design does not violate the temperature constraint.

\begin{table}[t]
    \centering
    \caption{Comparison of different DSE algorithms.}
    \begin{tabular}{lccc}
    \toprule
    \textbf{Method} & \textbf{Iterations to Converge} & \textbf{EDYP} & \textbf{Speedup} \\ 
    \midrule
     RL (PPO)     &  25,000 & 199.4 &  $1\times $  \\
     SA           &  3,180 & 153.5 & $7.9\times$   \\ 
     \textbf{ThermoDSE}    &  \textbf{850} & \textbf{132.4} &  $29.4\times$ \\ 
     \bottomrule
    \end{tabular}
    \label{tab:dse_comp}
    \vspace{-2mm}
\end{table}

\smallskip
Next, we evaluate the effect of the trust-region adjustment threshold (\textit{th} in Equation~\ref{eq_ltr}) on ThermoDSE’s convergence behavior. Fig.~\ref{fig_scbo_cove} shows the convergence results for three different threshold settings. When $th=4$, ThermoDSE converges quickly because a smaller threshold makes the adjustment more sensitive to failed exploration steps. It can achieve a feasible design within only 156 iterations; however, the resulting EDYP cost remains relatively high. When $th=16$, ThermoDSE identifies an almost optimal design with only 725 iterations. As the threshold increases, convergence becomes slower, but the observation cost decreases since the trust region is updated more gradually, allowing more tolerance for infeasible samples during boundary exploration.  

For constraint-aware chiplet-based design, the globally optimal design point typically lies extremely close to the temperature and area boundaries. Therefore, although a larger threshold requires more iterations to converge than smaller ones, it ultimately yields a better final design.

\smallskip
Finally, we compare ThermoDSE with SA and a PPO-based RL algorithm. To ensure fair comparison, both methods adopt the same architecture and task orchestration as ThermoDSE. The threshold of SCBO in ThermoDSE is fixed at 16. For RL, due to its extremely slow convergence, we allow up to 25K iterations. The reported EDYP value corresponds to the best design observed during RL exploration. Each method is executed twice, and Table~\ref{tab:dse_comp} lists the average convergence iterations and final EDYP results.  
ThermoDSE consistently finds a design point with considerably lower EDYP than both SA and RL. Moreover, ThermoDSE requires $3.7\times$ and $29.4\times$ fewer iterations than SA and RL, respectively, demonstrating that our algorithm is far more efficient and better suited for chiplet-based DNN accelerator design space exploration.

\begin{figure}[!t]
\vspace{-3mm}
\centering
    \subfloat[Area Breakdown]{\scriptsize \includegraphics[width=0.45 \linewidth]{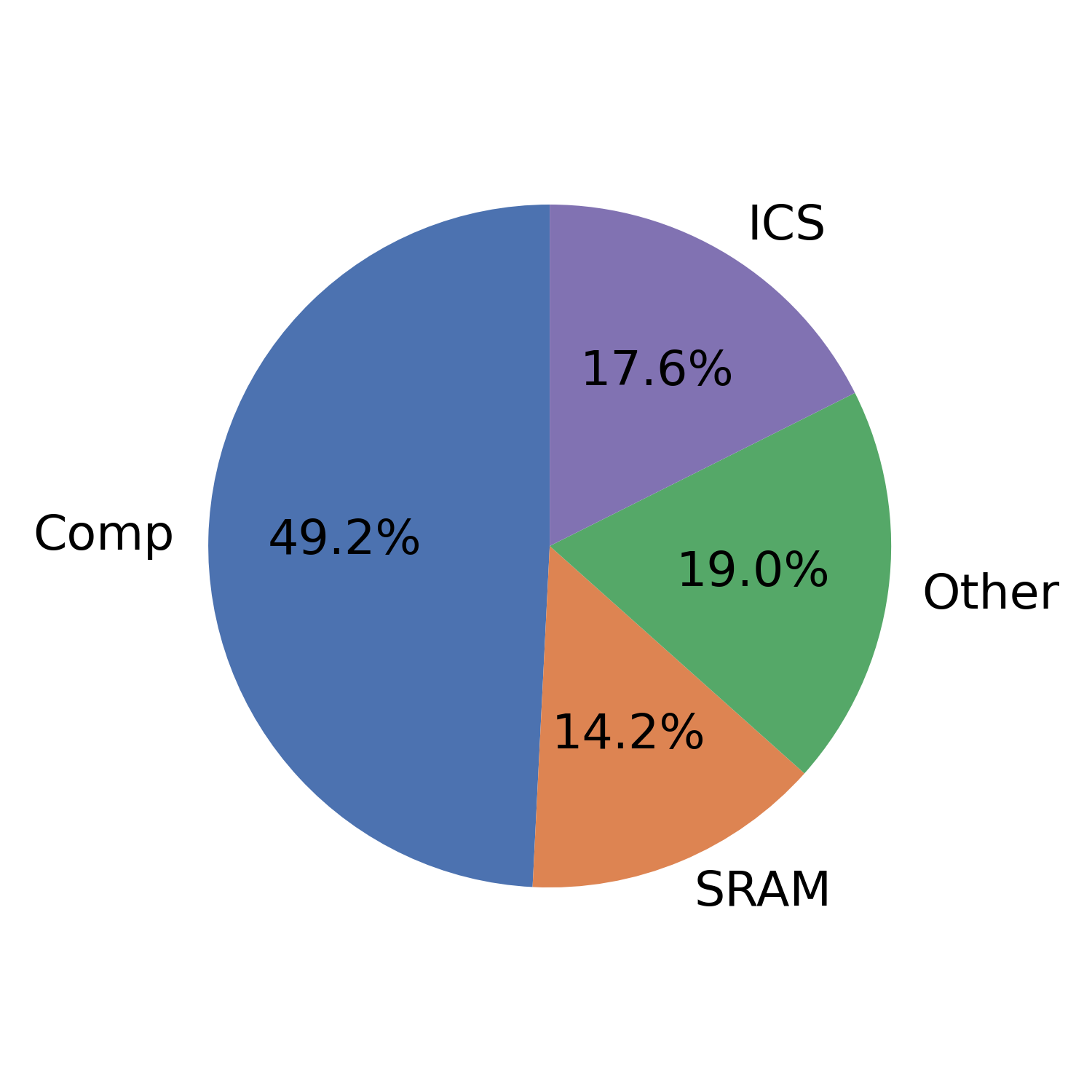}%
\label{fig_area_bkd}}
\hfil
\subfloat[Thermal Map]{\includegraphics[width=0.45 \linewidth]{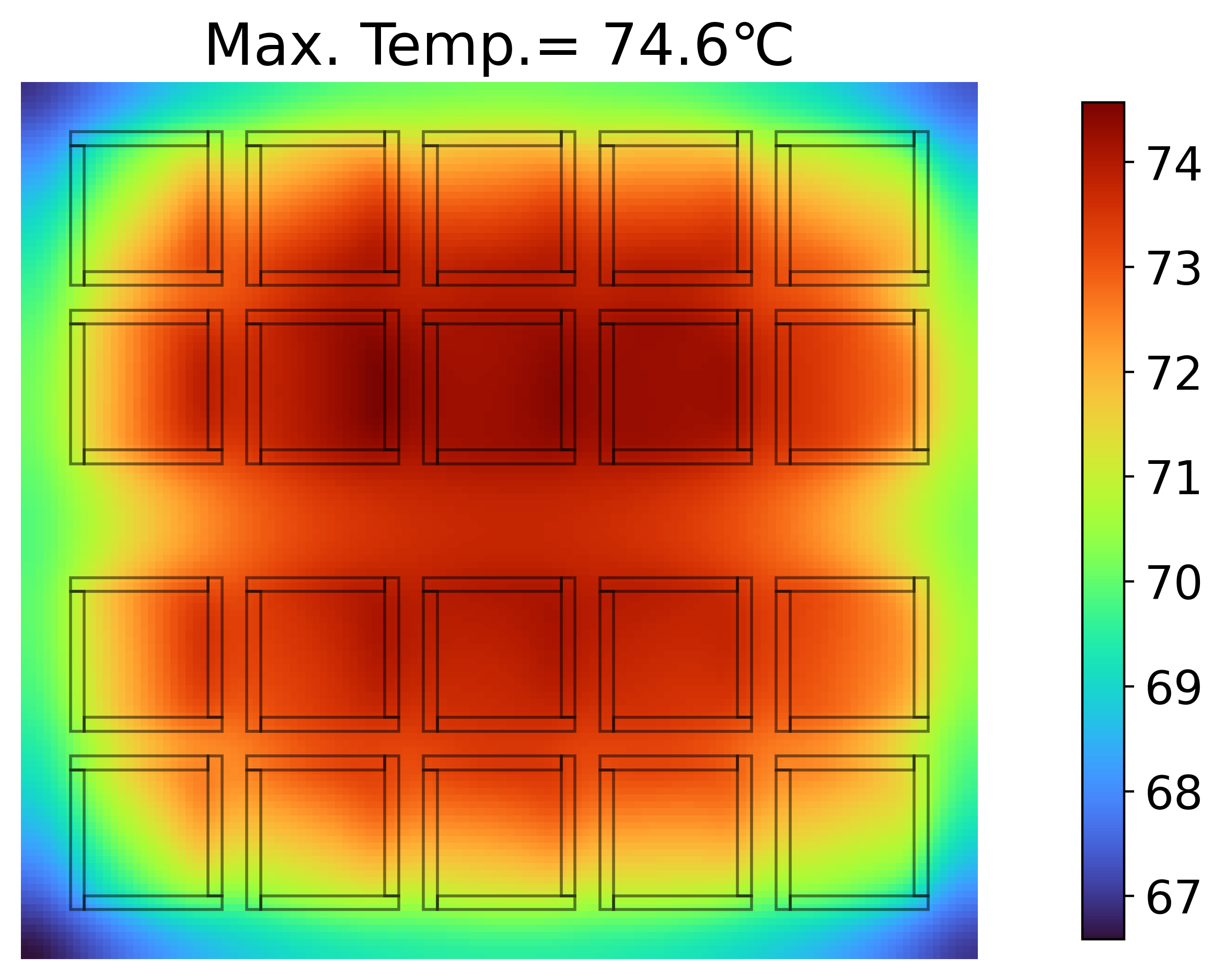}%
\label{fig_thermal_opt}}
\hfil
\caption{Analysis of area and thermal characteristics for the optimal design under $300\,mm^2$ area and $75\,^{\circ}$C constraints.} 
\label{fig_opt_info}
\end{figure}

\begin{figure}[!t]
    \centering
    \includegraphics[width=0.95\linewidth]{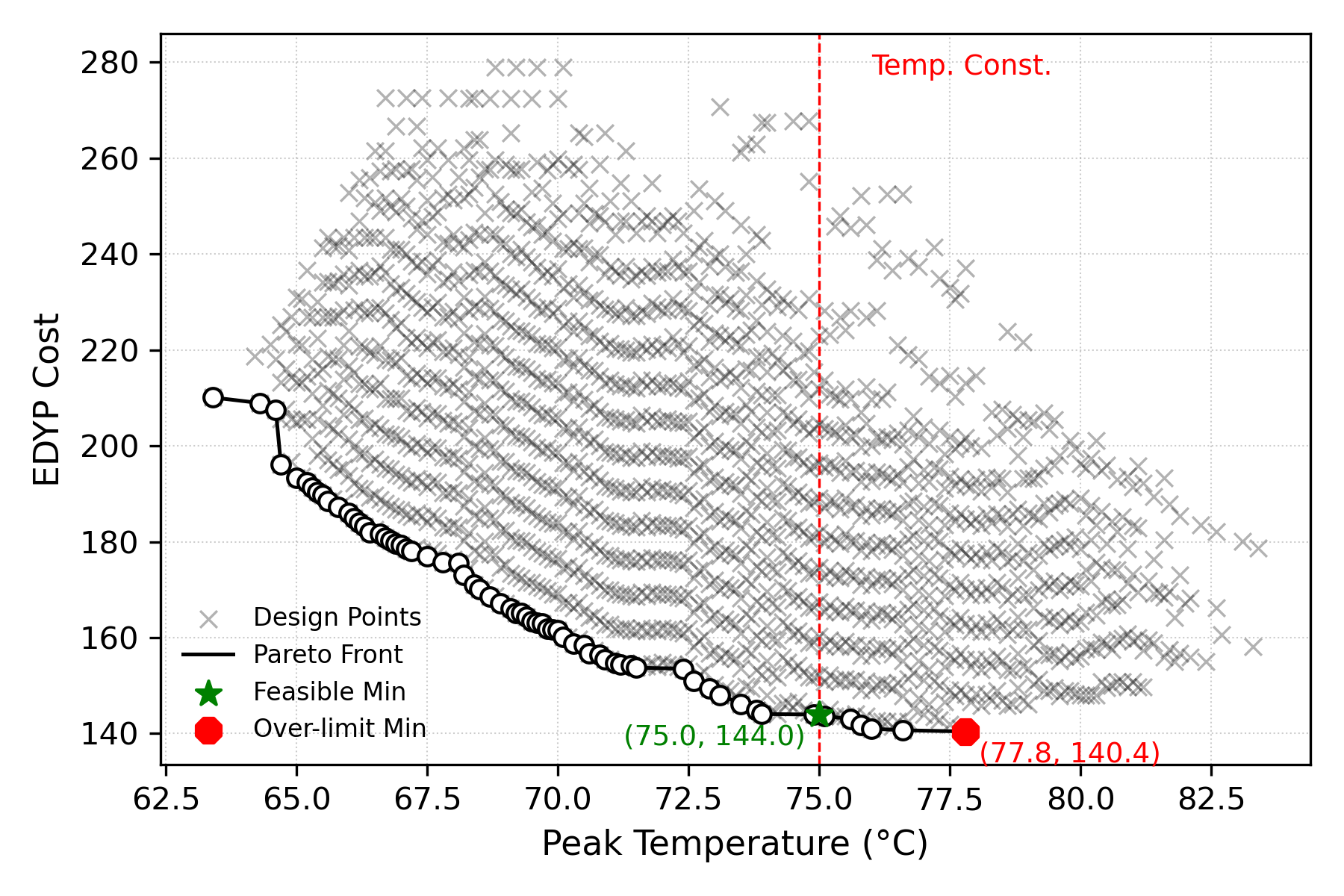}
    \vspace{-3mm}
    \caption{EDYP cost vs. peak temperature for chiplet-based DNN accelerators under $200\,mm^2$ area and $75\,^{\circ}$C thermal constraints. The feasible design (green) represents [4, 4, 1, 2, 2.3, 144, 112, 1024KB, 240], achieving a delay of $2.27\,ms$, energy consumption of $60\,mJ$, and a yield of 0.943. The cost-minimum design (red) corresponds to [5, 4, 1, 1, 0, 128, 112, 1024KB, 208], with a delay of $2.05\,ms$, energy consumption of $59.7\,mJ$, and a yield of 0.87.}
    \label{fig_parato_200}
    \vspace{-4mm}
\end{figure}
\subsection{Analysis of Performance, Area, and Temperature}
Our DSE framework identifies an almost optimal design point for chiplet-based DNN accelerators targeting AR/VR workloads.  
Fig.~\ref{fig_opt_info}(a) shows the area breakdown of the final design. As AI workloads are highly computation-intensive and data can be efficiently shared among NPU cores under fine-grained task modeling, the DSE algorithm allocates approximately 49\% of the total area to computation units, particularly to the matrix units. The on-chip buffer accounts for 14.2\% less area than the computation units. 

Compared with B2 and B3, the chiplet-based DNN accelerators with NoC/NoP allow more area to be dedicated to computation. Consequently, ThermoDSE achieves lower delay under compute-dominated conditions, which are typical in AI workloads. In addition, 17.6\% of the chip area is reserved for cooling structures.  
Fig.~\ref{fig_opt_info}(b) illustrates the thermal distribution of the optimal design obtained by ThermoDSE. Dividing the entire interposer into two dies not only minimizes the energy cost of D2D interconnects but also mitigates central thermal clustering, effectively reducing peak temperature.

\smallskip
We further explore the design space under a tighter area constraint of $200\,mm^2$. Fig.~\ref{fig_parato_200} presents the relationship of EDYP cost versus peak temperature. When the area budget becomes limited, the peak-temperature constraint noticeably restricts performance improvement due to the reduced thermal dissipation area available for cooling.  
Furthermore, the minimum achievable EDYP cost under a $300\,mm^2$ area budget is significantly lower than that of the $200\,mm^2$ configuration. These results highlight that the chiplet-based DNN accelerator design involves a complex trade-off among performance, area, and temperature.

\section{Conclusion and Future Work}
In this work, we presented \textbf{ThermoDSE}, a thermal-aware and comprehensive design space exploration framework for chiplet-based DNN accelerators. ThermoDSE integrates fine-grained task modeling with a uniform simulation and optimization framework that simultaneously considers chiplet granularity, NPU core granularity, task orchestration, and inter‑chiplet communication under strict thermal and area constraints.  
Compared with prior studies, our framework delivers significantly more accurate performance and thermal estimations. ThermoDSE achieves up to $3.5\times$ EDYP cost than state-of-the-art Simba-like designs and other optimization frameworks, respectively. Furthermore, the proposed DSE algorithm based on TED and SCBO attains superior design quality with $3.7\times$ and $29.4\times$ faster convergence compared with traditional SA and RL methods. In summary, ThermoDSE provides a scalable and physically-aware foundation for next-generation chiplet-based DNN accelerators.

\subsection{Key Takeaways}
The main insights obtained from this study are summarized as follows:
\begin{itemize}
    \item \textbf{Chiplet granularity matters.} Prior works often treated a single NPU core as one chiplet to maximize yield. However, overly fine-grained partitioning increases inter-chip communication energy and exacerbates thermal hot spots due to poor cooling on the bottom-layer interposer. Early-stage exploration of chiplet granularity is therefore essential, especially for energy- and thermally-constrained edge scenarios.
    \item \textbf{Integrated modeling is crucial.} Accurate co-simulation of data communication, computation, and fine-grained task orchestration is necessary. Ignoring these interactions can cause large deviations in performance and temperature estimation, leading to insufficient cooling design and incorrect performance budgeting.
    \item \textbf{Constraint-aware exploration is key.} The optimal design for chiplet-based accelerators often lies near constraint boundaries. Hence, efficient exploration methods such as our TED+SCBO algorithm are required. 
    %
\end{itemize}
\subsection{Future Work}
This study focuses on 2.5D chiplet integration. Although 3D stacking can offer higher integration density and improved performance, it also introduces much greater complexity in thermal management and design exploration. In future work, we plan to extend ThermoDSE to support 3D heterogeneous integration, enabling comprehensive co-optimization across performance, thermal, and reliability dimensions.



\vskip 0pt plus -1fil
\vspace*{-2\baselineskip}
\begin{IEEEbiography}
[{\includegraphics[width=1in,height=1.25in,clip,keepaspectratio]
{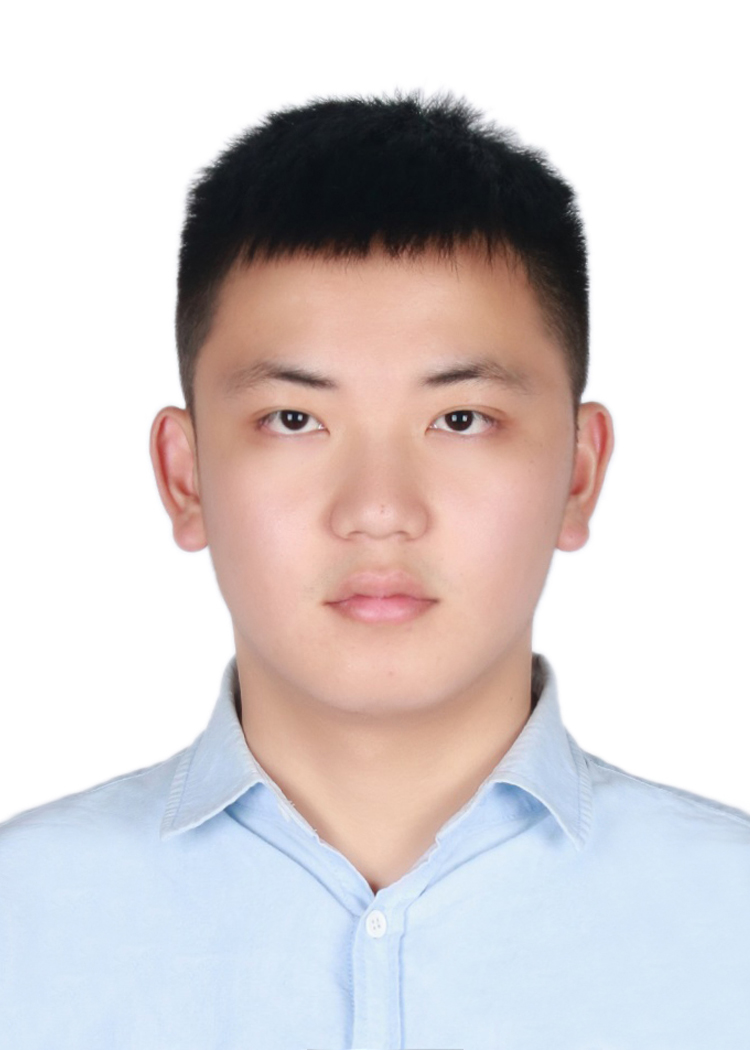}}]{Jian Peng} received his B.S. degree from University of Electronic Science and Technology of China, Chengdu, China, in 2019, and the MPhil degree from the Department of Electronic and Computer Engineering, The Hong Kong University of Science and Technology, Hong Kong, in 2022. He is currently a PhD student in the same department and university. His current research interests include power modeling, power and thermal management, and energy-efficient DNN accelerator designs.
\end{IEEEbiography}
\vskip 0pt plus -1fil
\vspace*{-2\baselineskip}
\begin{IEEEbiography}
[{\includegraphics[width=1in,height=1.25in,clip,keepaspectratio]
{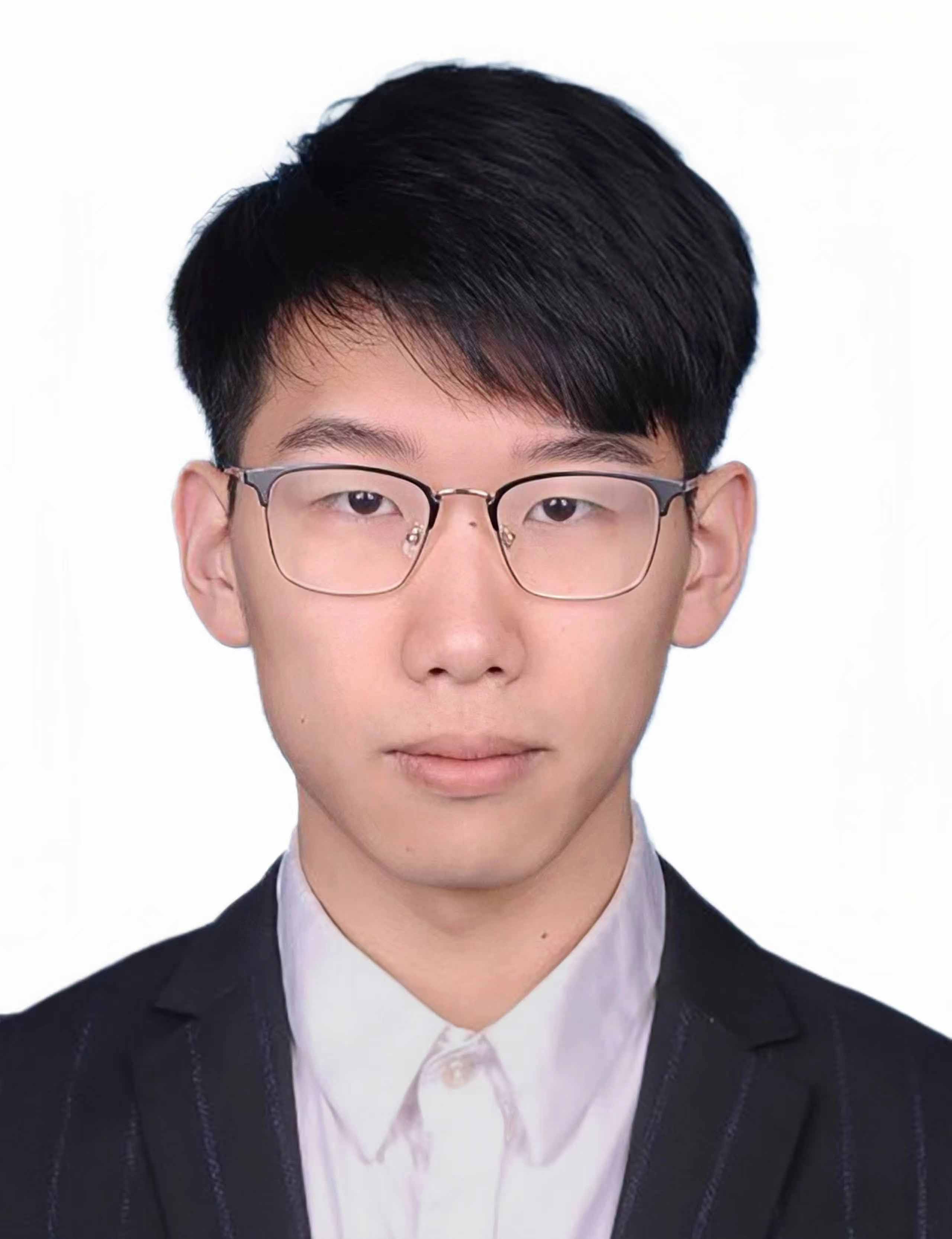}}]{Hanwei Fan} is currently a PhD candidate in Reconfigurable Computing System Laboratory (RCSL) at the Department of Electronic \& Computer Engineering in HKUST under the supervision of Professor Wei ZHANG. Previously, He obtained the BEng Degree in Automation from Beijing Institute of Technology and the MPhil Degree in Microelectronics from HKUST. His research interests include computer architecture, hardware-software co-design, and machine learning for EDA.
\end{IEEEbiography}
\vskip 0pt plus -1fil
\vspace*{-2\baselineskip}
\begin{IEEEbiography}
[{\includegraphics[width=1in,height=1.25in,clip,keepaspectratio]
{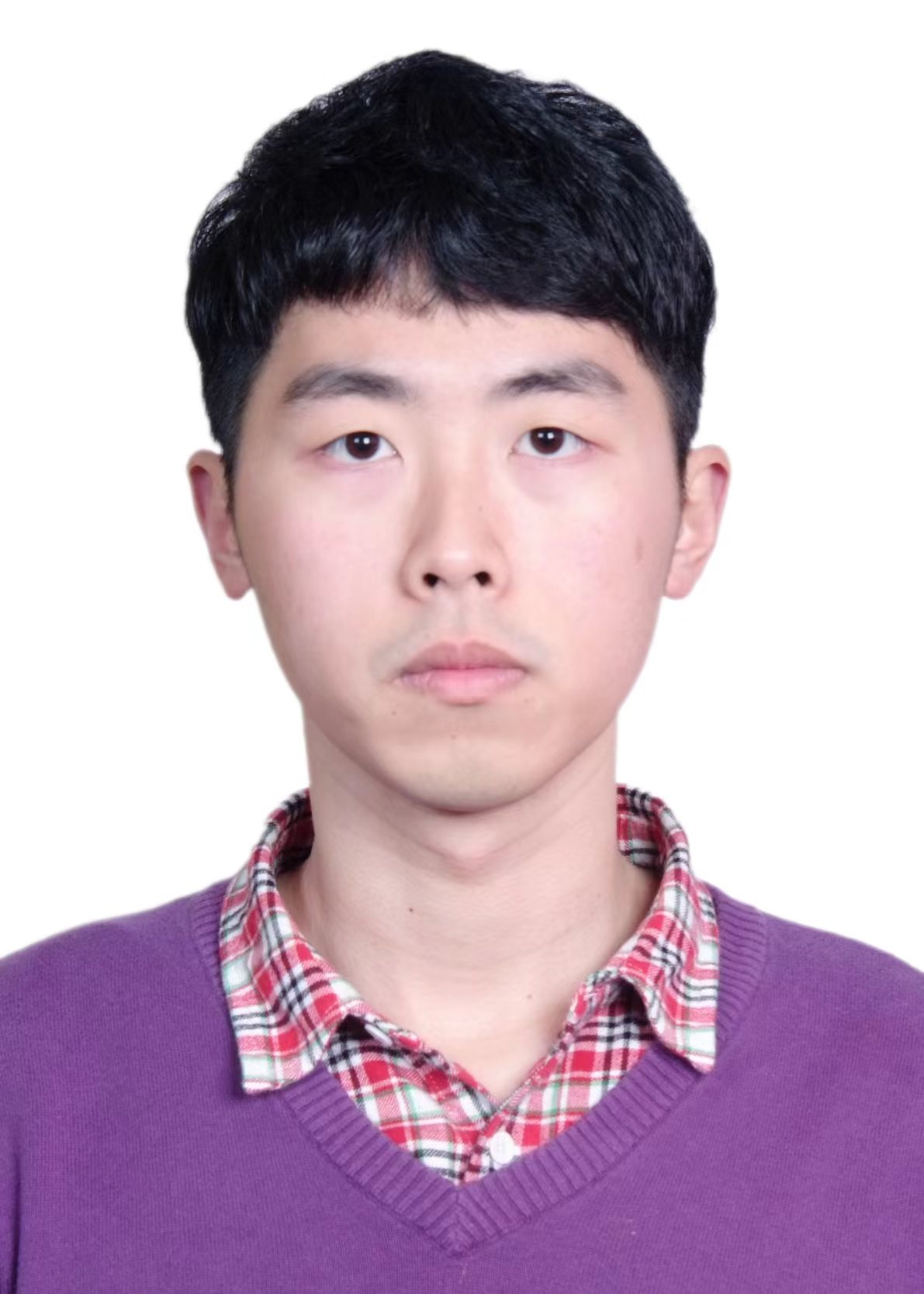}}]{Jingbo Jiang} received his B.S. degree from Hefei University of Technology, Hefei, China, in 2016, and his M.S. degree from the Hong Kong University of Science and Technology, Hong Kong, in 2020. He is currently working towards his Ph.D. degree at Hong
Kong University of Science and Technology, Hong Kong, China. His research interests are in circuit design for implementing deep learning algorithms.
\end{IEEEbiography}
\vskip 0pt plus -1fil
\vspace*{-2\baselineskip}
\begin{IEEEbiography}
[{\includegraphics[width=1in,height=1.25in,clip,keepaspectratio]
{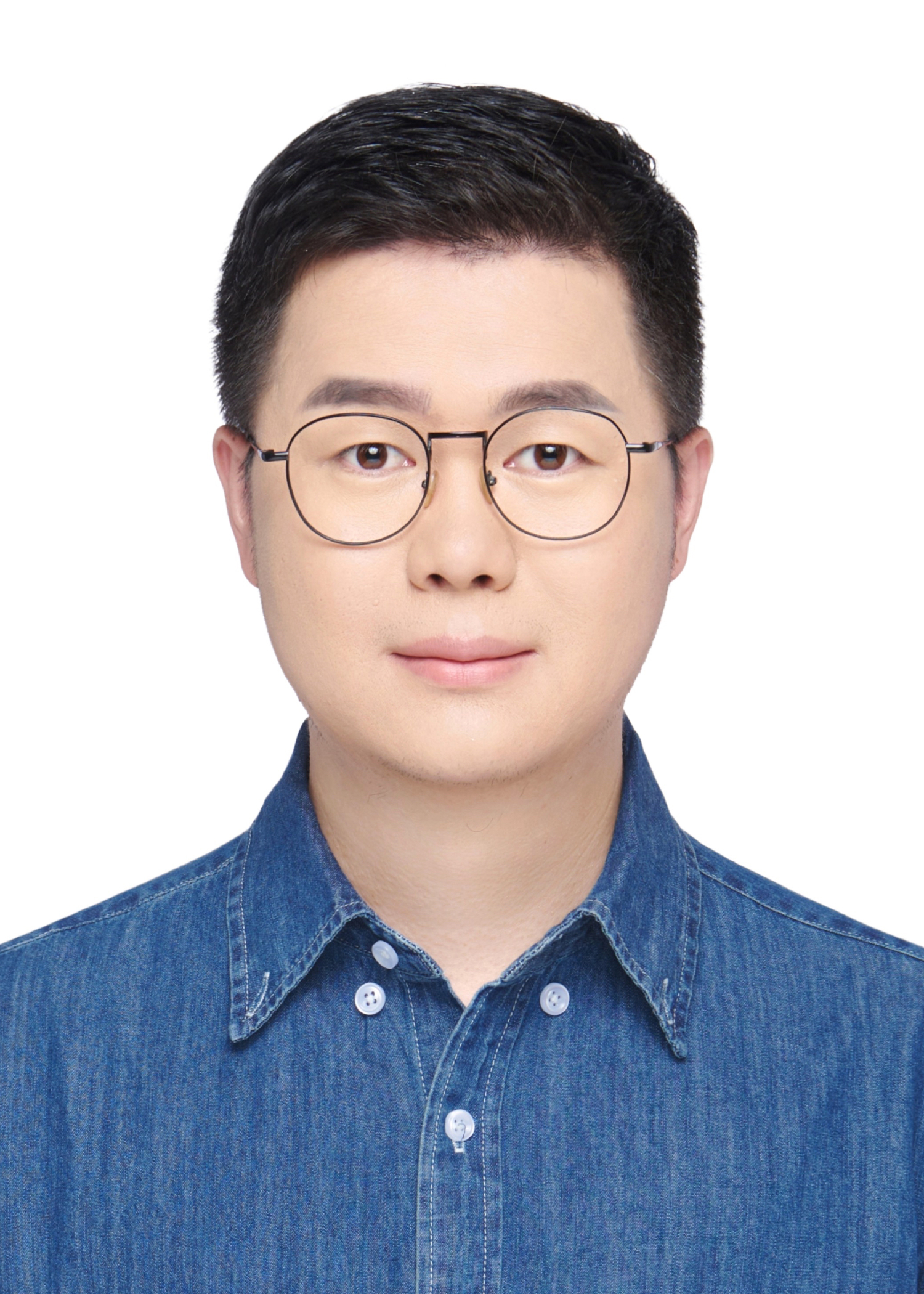}}] {Lin Jiang} is currently an associate professor at Northeastern University, China. He received his B.S. degree from Wuhan University and his M.S. degree from the University of Science and Technology of China. Dr. Jiang completed his Ph.D. at Clarkson University in New York, USA, and subsequently worked as a postdoctoral fellow at The Hong Kong University of Science and Technology. His research interests include thermal-aware physical design of processors, thermal management of 3D ICs, thermal modeling of multi-core CPUs and GPUs using data-driven approaches, reduced-order modeling, and finite element methods. He was a recipient of the Prof. Avram Bar-Cohen Best Paper Award from the $21^{st}$ Intersociety Conference on Thermal and Thermomechanical Phenomena in Electronic Systems.
\end{IEEEbiography}
\vskip 0pt plus -1fil
\vspace*{-2\baselineskip}
\begin{IEEEbiography}
[{\includegraphics[width=1in,height=1.25in, clip,keepaspectratio]{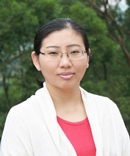}}]{Wei Zhang} (Fellow, IEEE) is a Professor at HKUST, where she established the RCSL. She received her PhD in Electrical Engineering from Princeton University, receiving the Wu Prize for research excellence. Prior to joining HKUST in 2013, she was an Assistant Professor at Nanyang Technological University, Singapore. She has authored over 150 peer-reviewed papers, winning Best Paper Awards at ISVLSI 2009, ICCAD 2017, and ICCAD 2022. Her research interests include reconfigurable systems, FPGA-based design, EDA, computer architecture, and embedded system security. Prof. Zhang serves as an Associate Editor for major journals including IEEE TCAD, IEEE TVLSI, and ACM TRETS. She has served as General Chair or Vice-Chair of ISVLSI 2018, FPT 2022 and ASAP 2024. She also serves on many organization committees and technical program committees, such as DAC, ICCAD, DATE, FPGA, etc. 
\end{IEEEbiography}
 




\vfill

\end{document}